\documentclass[man,11pt,letterpaper,hidelinks,floatsintext]{apa7} 

\usepackage{ragged2e}
\usepackage{indentfirst}
\usepackage{graphicx}
\usepackage{amsmath}
\usepackage{booktabs}
\usepackage{tabularx}

\usepackage{longtable}
\usepackage{array}
\usepackage{adjustbox}
\usepackage{hyperref}
\usepackage{orcidlink}
\usepackage[american]{babel}
\usepackage[utf8]{inputenc}    
\usepackage[T1]{fontenc}       
\usepackage{textcomp}          
\usepackage{csquotes} 
\usepackage{url} 
\usepackage{doi} 
\usepackage{eurosym}
\newcolumntype{s}{>{\setbox0=\hbox\bgroup}c<{\egroup}@{}} 
\usepackage{subcaption}
\usepackage{caption}
\usepackage{tikz}
\usetikzlibrary{decorations.pathreplacing,positioning,backgrounds, fit, calc}

\xdefinecolor{tuorange}{RGB}{227, 105, 19}
\xdefinecolor{tuyellow}{RGB}{242, 189, 0}
\xdefinecolor{tucitron}{RGB}{249, 219, 0}
\definecolor{gr}{rgb}{0.84, 0.89, 0.83}
\definecolor{softutgold}{RGB}{235, 217, 181} 
\definecolor{softutred}{RGB}{227, 170, 147} 
\definecolor{softutblue}{RGB}{162, 165, 230} 
\definecolor{softutgreen}{RGB}{198, 220, 207} 

\definecolor{utred}{RGB}{165,30,55} 
\definecolor{utred2}{RGB}{227,170,147}

\definecolor{utblue}{RGB}{50,65,75} 
\definecolor{utblue2}{RGB}{25,120,178} 
\definecolor{utblue3}{RGB}{105,178,205} 

\definecolor{utgreen}{RGB}{74,120,55} 
\definecolor{utgreen2}{RGB}{138,174,93} 
\definecolor{utgreen3}{RGB}{145,191,169} 

\definecolor{utgold}{RGB}{180,160,105}
\definecolor{utgold2}{RGB}{233,200,140}
\definecolor{utgold3}{RGB}{235,217,181}

\definecolor{utgrey}{RGB}{195,195,195} 
\definecolor{uthellgrey}{RGB}{203,203,203} 
\definecolor{utlightgrey}{RGB}{221,221,221}

\usepackage[style=apa, backend=biber]{biblatex}
\title{Cross-Course Generalizability of SRL-Aligned Predictive Models Using Digital Learning Traces}
\shorttitle{Cross-Course Generalizability}

\author{Jakob Schwerter$^{1,2}$\orcidlink{0000-0001-5818-2431}, 
    Loreen Sabel$^2$\orcidlink{0000-0002-9832-8842},
    Judith Bose$^2$, 
    Matthew L. Bernacki$^3$\orcidlink{0000-0003-1279-2829},
    Di Xu$^4$\orcidlink{0000-0002-8160-4783},
    Marko Schmellenkamp$^5$\orcidlink{0000-0003-3966-6590}, 
    Thomas Zeume$^5$\orcidlink{0000-0002-5186-7507},
    Philipp Doebler$^2$\orcidlink{0000-0002-2946-8526} 
}
\vspace{6pt}
\authorsaffiliations{
{$^1$ Hector Research Institute of Education Sciences and Psychology, University of Tübingen}\\
{$^2$ TU Dortmund University}\\
{$^3$ University of North-Carolina, Chapel Hill}\\
{$^4$ University of California, Irvine}\\
{$^5$ Ruhr University Bochum}
}

\authornote{
Correspondence concerning this manuscript should be directed to Jakob Schwerter, University of Tübingen, Hector Research Institute of Education Sciences and Psychology (HIB), Europastr. 6, 72072 Tübingen, Germany. Email: \href{mailto:jakob.schwerter@uni-tuebingen.de}{jakob.schwerter@uni-tuebingen.de}

\addORCIDlink{Jakob Schwerter}{0000-0001-5818-2431}

\addORCIDlink{Loreen Sabel}{0000-0002-9832-8842}

\addORCIDlink{Matthew L. Bernacki}{0000-0003-1279-2829}

\addORCIDlink{Di Xu}{0000-0002-8160-4783}

\addORCIDlink{Marko Schmellenkamp}{0000-0003-3966-6590}

\addORCIDlink{Thomas Zeume}{0000-0002-5186-7507}

\addORCIDlink{Philipp Doebler}{0000-0002-2946-8526}

\medskip

\textbf{Funding}: This work was supported by ``From Prediction to Agile Interventions in the Social Sciences (FAIR, PROFILNRW-2020-068)'', by Deutsche Forschungsgemeinschaft (DFG, German Research Foundation), grant 448468041 on ``Formal Foundations for Teaching Support Systems in Theoretical Computer Science'', and by Intramural Funding of the ``Trustworthy Data Science and Security'' at TU Dortmund. The sole responsibility for the content of this publication lies with the authors.

\textbf{Declarations}: All authors certify that they have no affiliations with or involvement in any organization or entity with any financial or non-financial interest in the subject matter or materials discussed in this manuscript. The authors have no financial or proprietary interests in any material discussed in this article.

\textbf{Declaration of generative AI and AI-assisted technologies in the writing process}: During the preparation of this work, the authors used GPT-5 and DeepL for language-focused editing of human-written text. After using these tools/services, the authors reviewed and edited the content as needed and take full responsibility for the content of the publication.

\textbf{Conflict of Interest}: The authors have no conflicts of interest to declare.

\textbf{Ethical Approval}: Approval was granted by the Ethics Committee of TU Dortmund University.
}

\abstract{
STEM dropout rates remain high at universities, particularly in computer science programs with theory-intensive courses. Digital learning environments now capture rich behavioral data that could help identify struggling students early, yet the generalizability of data-driven prediction models across courses and institutions remains uncertain. Guided by self-regulated learning (SRL) theory, this study analyzed multimodal digital-trace data from three undergraduate theoretical computer science courses ($N_1 = 137$, $N_2 = 104$, $N_3 = 148$) at two universities. Weekly SRL-aligned digital-trace indicators were modeled using Elastic Net, Random Forest, and XGBoost to evaluate predictive performance over time and across settings, and model calibration both within and across courses. Early prediction of at-risk students was feasible, with SRL-related behaviors such as time management, effort regulation, and sustained engagement emerging as key predictors. While Random Forest achieved the highest in-sample accuracy, Elastic Net generalized more robustly across contexts. Out-of-sample accuracy and calibration declined between institutions with different base rates, underscoring the contextual nature of predictive analytics in higher education. These findings suggest that digital learning traces enable early identification of at-risk students within courses, but generalizing predictive models beyond their original context requires caution, particularly if the at-risk rates differ between contexts.
}


\begin{document}
\maketitle
\justifying
\setlength{\parindent}{1.27cm}


Higher education faces declining enrollment and growing dropout concerns, while also grappling with a student population that is becoming increasingly diverse in academic preparation and background \parencite{HeubleinEA2022, NCES2023, OECD2019}. These challenges are particularly pronounced in science, technology, engineering, and mathematics (STEM) disciplines, where maintaining persistence through degree completion has proven especially difficult. Across OECD countries, STEM completion rates average roughly 68 percent---well below those in health, welfare, or social sciences programs \parencite{oecd_education_2025}. Programs in information and communication technology and engineering disciplines consistently exhibit the lowest completion rates, indicating that this pattern reflects a systemic rather than institution-specific issue. One commonly cited reason is that students in these fields encounter theory-intensive coursework early in their studies, and many students struggle to manage the substantial workload and sustain effective study habits \parencite{SollEA2023}. When these difficulties accumulate \parencite{BrownEA2022}, withdrawal becomes more likely, which leads to important consequences for students, institutions, and society \parencite{SchneiderBB2019, FaasEA2018}.

These challenges have prompted instructors and institutions to look for ways to better support students’ day-to-day learning processes, particularly in the large and demanding courses where many students struggle. One promising approach stems from the widespread integration of digital learning environments into college instruction. These environments structure opportunities for regular practice, provide immediate feedback, and make students’ engagement patterns visible in ways that traditional settings do not \parencite{HellasEA2018}. A growing number of studies consistently show that active participation in digital learning platforms is associated with better performance and persistence rates across a range of higher education settings, including, for example business-administration courses \parencite{SchwerterDBM2022, ForsterWM2018}, social-science statistics courses \parencite{SchwerterB2024}, psychology statistics courses \parencite{janson_influence_2024, Janson.etal2024}, and undergraduate science courses \parencite{Mefferd.Bernacki2023}, but evidence remains sparse for computer science—despite high dropout rates and distinct cognitive demands in theory-driven coursework. These findings suggest that digital trace data capture meaningful aspects of how students navigate complex learning tasks, enabling early-warning models that flag risk weeks before exam outcomes are available \parencite{BernackiEA2020, TsaiEA2020, vonKeyserlingkEA2023}. %

Self-regulated learning \parencite[SRL; see][]{EfklidesS2024, greene_self-regulation_2024} offers a robust theoretical framework for interpreting these digital traces and explaining why early learning behaviors predict later achievement \parencite{BernackiEA2020}. SRL describes learners’ ability to direct their cognition, motivation, and behavior toward achieving academic goals. As students transition into college-level STEM coursework, these regulatory processes become particularly consequential, because students must manage demanding tasks with less external structure than they experienced in secondary school \parencite{FaasEA2018, WoltersB2021}. 

Predictive learning analytics translates these SRL-aligned digital traces into early-warning models intended to identify students who may be at risk of failure \parencite{sghir_recent_2023}. A growing body of research demonstrates that exam performance can often be predicted within the first weeks of a semester \parencite{ArizmendiEA2023, BernackiEA2025, CoglianoEA2022}. When accurate, these early predictions create opportunities for instructors to reach out to struggling students, offer timely feedback, or connect them with additional resources before disengagement becomes more difficult to address \parencite{CoglianoEA2022}. As a result, predictive analytics and early-warning systems have been adopted widely across higher education. In the United States, roughly one-third of higher-education institutions---approximately 1,400 colleges and universities---now use early-warning or predictive analytics systems, representing a market exceeding \$500 million \parencite{barshay_colleges_2019}. In Germany the state of North Rhine-Westphalia has committed nearly \euro{4} million (2024–2026) to support the development of AI-based learning analytics initiatives \parencite{mkw_2023}. 

Despite this momentum, however, important questions remain about the conditions under which prediction models provide reliable information and the extent to which they can be used beyond the specific courses or institutions in which they were developed. On the one hand, the rapid adoption of early-warning systems across institutions underscores the importance of generalizability, since colleges often hope to reuse models across semesters or adapt them to similar courses instead of rebuilding them each time \parencite{mathrani_perspectives_2021}. On the other hand, theoretical considerations suggest that generalizable predictive models require parsimonious feature sets that are broadly applicable across settings. In practice, however, many high performing models rely on behaviors that are tightly coupled to the structure of a particular course, which limits robustness when learning conditions change \parencite{ArizmendiEA2023}. The mapping of low-level digital events onto SRL processes also varies by platform and instructional design, constraining portability unless features capture theory-aligned constructs such as planning, effort regulation, or pacing \parencite{DuEA2023}.  Although early research demonstrates promising reapplication within single contexts \parencite[e.g.,][]{plumley_codesigning_2024}, cross-context generalization remains rare despite conceptual calls to treat generalization as a core criterion for trustworthy learning analytics \parencite{mathrani_perspectives_2021}.

Beyond generalizability, most studies emphasize discrimination (e.g., AUC) but rarely test calibration—whether predicted probabilities match observed risk—yet calibration is central for interpretable, threshold-based support and equitable resource allocation. Calibration ensures that estimated probabilities correspond to actual outcomes, enabling threshold-based interventions and resource allocation that are both equitable and interpretable \parencite{PlattEA1999, PhelpsEA2024, DeCockCampo2025}. Attention to calibration is therefore essential for improving fairness, interpretability, and the practical utility of predictive learning analytics \parencite{mathrani_perspectives_2021}. This study addresses these gaps by examining predictive performance, generalizability, and calibration across three theory intensive computer science courses at two universities. 
By combining cross course and cross institution validation analysis with attention to calibration, the study provides empirical evidence that is currently missing from the literature and clarifies the conditions under which early warning models can support scalable, meaningful guidance for students and instructors.

\section{Theoretical background and related work}


\subsection{SRL and digital traces}

SRL provides a theoretical foundation for understanding why students’ early engagement patterns in digital learning environments can predict later achievement, framing learning as an iterative cycle of planning, strategy use, monitoring, adaptation, and self-evaluation \parencite{EfklidesS2024, greene_self-regulation_2024}. In technology-enhanced courses, elements of these cycles manifest as behavioral traces in learning management systems, allowing low-level digital events (e.g., time allocation, task completion) to be aligned with higher-level SRL processes such as time management, effort regulation, and strategic revision \parencite{MatchaEA2020, DuEA2023}.

Because trace data are collected continuously and at scale in learning management systems (LMS), predictive learning analytics based on these traces have become widely used to identify and support at-risk students in higher education \parencite{ArizmendiEA2023, sghir_recent_2023}. Many studies operationalize behavioral indicators, such as logins, submissions, and page views, as indirect evidence of SRL, mapping these micro-level actions to macro-level processes of planning, monitoring, and regulation \parencite{MatchaEA2020, DuEA2023}. The interpretability of such mappings depends on theory-aligned feature construction and consistent instrumentation, motivating calls for explicit theory-to-feature documentation to strengthen construct validity and generalizability \parencite{CukurovaEA2020, DuEA2023}. 

Recent multimodal studies have strengthened this theoretical integration by synchronizing think-aloud protocols with timestamped digital traces, showing that clusters of actions correspond to SRL macroprocesses and supporting the construct and predictive validity of SRL-aligned features \parencite{BernackiEA2025, Yu_inpress_SRLSequences, Fan2022_ValiditySRL}. At the same time, not all SRL processes are equally observable in digital logs, as instrumentation and task design constrain what is captured, individual events may reflect multiple macroprocesses, and metacognitive judgments such as monitoring accuracy remain difficult to detect without additional modalities or metadata  \parencite{BernackiEA2025}. Although multimodal learning analytics can address some of these limitations by integrating complementary data sources (e.g., physiological, video, or verbal data), their methodological, practical, and ethical complexity limits widespread adoption \parencite{CukurovaEA2020, BernackiEA2020}. Consequently, LMS traces remain the most feasible and widely used data source for predictive learning analytics in higher education.

\subsection{Predictive learning analytics}

Predictive learning analytics encompasses a broad spectrum of analytical methods, ranging from interpretable regression models to complex machine-learning algorithms. Tree-based approaches such as Random Forests and Extreme Gradient Boosting often yield higher predictive accuracy, whereas regularized regressions such as Elastic Net balance performance with interpretability and parsimony \parencite{ArizmendiEA2023}.

Building on these methodological advances, empirical studies consistently show that exam performance can be predicted within the first weeks of a semester using digital trace data, providing educators with greater lead time to implement targeted support \parencite{BernackiEA2020}. For instance, in STEM lecture contexts, failure in end-of-term exams has been predicted with practical accuracy of 70–80 percent within the first few weeks, enabling proactive interventions that support students’ self-regulation and improve later outcomes \parencite{ArizmendiEA2023, CoglianoEA2022, BernackiEA2020}.

However, large-scale findings indicate that predictive accuracy depends on data type and context: administrative records predict outcomes well for returning students, whereas behavioral traces from LMS offer the greatest added value for new students without prior academic histories \parencite{bird_is_2025}. Efforts to increase predictive accuracy by adding demographic variables raise important ethical and methodological concerns. While demographic variables such as gender, ethnicity, and prior GPA can enhance predictive performance or serve as statistical controls, their use risks reinforcing inequities, reducing transparency, undermining trust in analytic systems, and contributing little to the design of actionable instructional supports \parencite{ArizmendiEA2023, BernackiEA2020, ferguson2017evidence, slade2013learning}. 
Consequently, recent work encourages prioritizing SRL-aligned behavioral indicators, documenting modeling decisions transparently, and assessing differential model validity across student subgroups to ensure equitable interpretations.

A central challenge for cross-course generalizability is that digital traces are not context-free indicators of learning. The behaviors recorded in LMS logs are shaped not only by students’ self-regulation, but also by how instructors design the course, which tools are activated, how activities are sequenced, and whether engagement is incentivized or required. \textcite{Park.etal2016}, for example, showed that courses cluster into distinct LMS-use profiles, indicating that the same platform can support substantially different forms of activity across courses. This implies that a given trace, such as repeated page views, early submissions, or forum activity, may reflect different underlying processes depending on the instructional design in which it occurs. In this sense, trace data are jointly produced by learner behavior and course architecture, which complicates efforts to transfer predictive models across settings.

Prior predictive-analytics research supports this concern. \textcite{conijn_predicting_2017} compared 17 blended courses within a single institution and found that predictive results varied strongly across courses, with low portability of prediction models and no single set of LMS variables that consistently predicted performance across contexts. They argued that differences in course type, learning design, and available LMS modules likely contributed to this instability. From an SRL-measurement perspective, more recent work suggests that generalizability depends not only on model transfer but also on the validity of the trace-to-construct mapping. \textcite{BernackiEA2025} demonstrated that some digital events can be validated as traces of SRL and can predict achievement across authentic course settings and a subsequent semester, but they also showed that some events map onto multiple SRL macroprocesses. Likewise, \textcite{Song.etal2025b} argued that trace-SRL research faces a validity–generalisability trade-off and reported that universal trace indicators did not transfer cleanly across multidisciplinary subjects, with engagement proving easier to detect than planning or reflection. Taken together, these studies suggest that cross-course generalizability is most plausible when models rely on theoretically interpretable families of behaviors and when the target courses share sufficient similarity in design, instrumentation, and incentive structure.

\textcite{bernacki_clickstream_2026} have more recently proposed an approach to decompose a course’s instructional design into component parts that capture this context in ways that can further promote generalizability of analytic approaches based on course features. Their taxonomy involves the identification of content units guided by instructional objectives appraised by course assessments, and populated with learning materials that align to learning objectives. They further apply temporal tagging from course calendars and learning theory tags that describe the cognitive and metacognitive processes learners might be able to undertake with a similar resource provided within a course that affords instructional designs and learning conditions. This decomposition and coding lays groundwork for us to consider how such conditions are sustained or vary across courses and across institutions that offer them.

\subsection{Methodological challenges and transferability}

Despite promising within-course prediction results, cross-course generalizability remains a central unresolved problem in trace-based learning analytics because the meaning, availability, and incentive structure of digital behaviors often vary with course design.
Although SRL-aligned features enhance theoretical coherence, the behavioral traces underlying them often vary with course design and assessment structure. For instance, a behavior pattern that signals proactive regulation in one course may reflect mere compliance in another, depending on how engagement is incentivized \parencite{jovanovic_students_2021, CukurovaEA2020}. Systematic reviews therefore caution that even high-performing models may remain context-bound without explicit cross-course or cross-institution validation \parencite{mathrani_perspectives_2021, sghir_recent_2023}. 

A second methodological concern is related to probability calibration. Although most studies report discrimination metrics to show how well models separate higher- and lower-risk students, far fewer assess whether predicted probabilities accurately represent actual outcomes \parencite{TsaiEA2020, sghir_recent_2023}. 
Because miscalibration can lead to inefficient or inequitable allocation of support—particularly when base rates differ across contexts—calibration is essential for interpretable and practically useful early-warning systems \parencite{PlattEA1999, PhelpsEA2024, DeCockCampo2025}. Improving calibration therefore enhances both the interpretability and practical usefulness of predictive learning analytics.

\subsection{Present study}

This study addresses these gaps by examining both the generalizability and calibration of SRL-aligned predictive models across courses with structurally similar learning environments. Using digital trace data from three theoretical computer science courses offered at two universities, we investigate how well models trained in one context perform when applied to another and whether the probabilities they generate offer reliable estimates of student risk.

We explore two sets of research questions. The first set of research questions focuses on model generalizability and feature transferability, examining the predictive performance of SRL-related indicators both within and across courses and institutions. Specifically, we assess (RQ1a) how early in the semester at-risk students can be identified from digital trace data, (RQ1b) whether predictive performance generalizes across courses and institutions, (RQ1c) which SRL-related indicators serve as the most informative predictors, and (RQ1d) the extent to which these predictive indicators overlap across contexts. The second main research question (RQ2) examines whether post hoc probability calibration improves the accuracy and interpretability of risk estimates, thereby supporting timely and targeted learning interventions.

By evaluating model performance across multiple courses and institutions, and testing the added value of probability calibration, this study provides first empirical evidence on when SRL-aligned predictions can be expected to generalize and how calibrated probabilities improve the accuracy of risk estimates. These insights offer concrete guidance for developing early-warning systems that remain informative and actionable across varied higher-education settings.

\section{Methods}
\subsection{Participants and Context}

The dataset for this study was drawn from the [anonymized] project, collected during the 2023–2024 winter term and included undergraduate students enrolled in three computer science courses offered at two German universities within the same metropolitan region. At University A, participants were enrolled in the course ``Theoretical Computer Science'' (TCS-A) and ``Logic'' (Logic-A). At University B, participants were enrolled in a parallel ``Logic'' (Logic-B) course.

All courses are delivered in person and consisted of a lecture covering core content and small-group exercise sessions of 20–30 students. The course on theoretical computer science (TCS-A) featured two 90-minute lectures and one 90-minute exercise session per week; the two logic courses (Logic-A and Logic-B) featured one 90-minute lecture per week and one 90-minute exercise session every other week. Additionally, the Logic-B course offered a weekly tutorial in which sample solutions to lecture-related assignments were discussed. Attendance at lectures, exercise sessions, and tutorials was voluntary. As depicted in Figure~\ref{fig:overlap}, courses TCS-A and Logic-A share an institutional context, whereas courses TCS-A and TCS-B share topical content. This structure provides a systematic basis for examining both intra- and inter-institutional generalizability of predictive models.

\begin{figure}
    \centering
    \caption{Graphical overview of course overlaps}
    \label{fig:overlap}
    \begin{subfigure}[b]{0.4\textwidth}
\begin{tikzpicture}[scale=0.65, transform shape]

\tikzset{
  course/.style={draw, rounded corners=4pt, inner sep=2pt,
                 minimum width=2.4cm, minimum height=0.9cm, fill=none},
  boxLeft/.style ={rounded corners=6pt, fill=utblue2, fill opacity=.5, 
                   inner sep=8pt,
                   label={[anchor=north ,xshift=4pt]north :{\bfseries\small\centering Same university}}},
 boxLeftPhantom/.style ={rounded corners=6pt, fill=utblue2, fill opacity=.0, 
                   inner sep=8pt,
                   label={[anchor=north ,xshift=4pt]north :{\bfseries\small\centering \phantom{Same university}}}},                  
  boxRight/.style={rounded corners=6pt, fill=utred2, fill opacity=.75, 
                   inner sep=8pt,
                   label={[anchor=north ,xshift=-4pt]north :{\bfseries\small\centering Same topic/course}}},
  boxRightPhantom/.style={rounded corners=6pt, fill=utred2, fill opacity=.0, 
                   inner sep=8pt,
                   label={[anchor=north ,xshift=-4pt]north :{\bfseries\small\centering \phantom{Same topic/course}}}},                 
  container/.style={draw=black, rounded corners=8pt, thick, fill=none, inner sep=10pt,
                   label={[anchor=north,yshift=-2pt]north:
                     }}
}

\node[course] (c1) at (-3.5,0) {Course TCS-A};
\node[course] (c2) at ( 0.0,0) {Course Logic-A};
\node[course] (c3) at ( 3.5,0) {Course Logic-B};

\begin{pgfonlayer}{background}
  \node[boxRightPhantom, fit=(c2)(c3)] (rightbox) {\phantom{Nothing}};
  \node[boxLeft,  fit=(c1)(c2)] (leftbox)  {};
\end{pgfonlayer}

\begin{pgfonlayer}{background}
  \node[container, fit=(leftbox)(rightbox)(c1)(c2)(c3), yshift=0mm] (big) {};
\end{pgfonlayer}
\end{tikzpicture}
\end{subfigure}
\qquad \quad
\begin{subfigure}[b]{0.4\textwidth}

\begin{tikzpicture}[scale=0.65, transform shape]

\tikzset{
  course/.style={draw, rounded corners=4pt, inner sep=2pt,
                 minimum width=2.4cm, minimum height=0.9cm, fill=none},
  boxLeft/.style ={rounded corners=6pt, fill=utblue2, fill opacity=.5, 
                   inner sep=8pt,
                   label={[anchor=north ,xshift=0pt]north :{\bfseries\small\centering Same university}}},
 boxLeftPhantom/.style ={rounded corners=6pt, fill=utblue2, fill opacity=.0, 
                   inner sep=8pt,
                   label={[anchor=north ,xshift=0pt]north :{\bfseries\small\centering \phantom{Same university}}}},                  
  boxRight/.style={rounded corners=6pt, fill=utred2, fill opacity=.75, 
                   inner sep=8pt,
                   label={[anchor=north ,xshift=-0pt]north :{\bfseries\small\centering Same topic/course}}},
  boxRightPhantom/.style={rounded corners=6pt, fill=utred2, fill opacity=.0, 
                   inner sep=8pt,
                   label={[anchor=north ,xshift=-0pt]north :{\bfseries\small\centering \phantom{Same topic/course}}}},                 
  container/.style={draw=black, rounded corners=8pt, thick, fill=none, inner sep=10pt,
                   label={[anchor=north,yshift=-2pt]north:
                     }}
}

\node[course] (c1) at (-3.5,0) {Course TCS-A};
\node[course] (c2) at ( 0,0) {Course Logic-A};
\node[course] (c3) at ( 3.5,0) {Course Logic-B};

\begin{pgfonlayer}{background}
  \node[boxRight, fit=(c2)(c3)] (rightbox) {\phantom{Nothing}};
  \node[boxLeftPhantom,  fit=(c1)(c2)] (leftbox)  {};
\end{pgfonlayer}

\begin{pgfonlayer}{background}
  \node[container, fit=(leftbox)(rightbox)(c1)(c2)(c3), yshift=0mm] (big) {};
\end{pgfonlayer}

\end{tikzpicture}
\end{subfigure}
  \begin{flushleft}
\footnotesize \textit{Note.} The diagram illustrates the structural relationships between the three analyzed courses. Courses TCS-A and Logic-A share the same institutional context (University A), whereas Logic-A and Logic-B share topical content. This structure allows examination of both within-institution and cross-institution model generalization.
\end{flushleft}
\end{figure}

A total of 389 undergraduate students participated in the study, including  137 students in Course TCS-A (20\% female, 26\% first-generation students), 104 students in Course Logic-A (25\% female, 32\% first-generation students), and 148 students in Course Logic-B (19\% female, 32\% first-generation students). The mean age across courses was 22 years. For more than 60\% of students, German was the language spoken at home (Course TCS-A: 63\%, Course Logic-A: 67\%, Course Logic-B: 62\%).

In University~A, the participating students either study computer science (Course~TCS-A: 35\%, Course~Logic-A: 41\%), applied computer science (TCS-A: 29\%, Logic-A: 41\%), IT security (TCS-A: 32\%, Logic-A: 18\%), and other (TCS-A: 4\%, Logic-A: 0\%). In University~B (Course~Logic-B), the majority of participating students study computer science (88\%) and a on few studied applied computer science (6\%) or another area of study (6\%).

At both universities, students could take the exam on two different dates. The first exam was held immediately after the lecture period, and the second took place just before the beginning of the following semester. Students who failed or (at Institution A) wished to improve their grade were permitted to retake the exam at the second date. Some students registered only for the second opportunity. Across the first exam dates, 197 students in Course TCS-A, 143 in Course Logic-A, and 193 in Course Logic-B took the exam, with participation rates ranging from 51\% to 59\%. At the second exam date, 77 to 106 students took the exam, and 40\% to 45\% of them participated in the study.


\subsubsection{Ethical Considerations}

The study was conducted in accordance with institutional and national ethical guidelines for educational research. Participation was voluntary, and informed consent was obtained from the students. The research protocol was reviewed and approved by an institutional review board (approval number anonymized for review). All data were anonymized prior to analysis to ensure confidentiality and compliance with data protection regulations.

\subsection{Learning Environment and Data Sources}

All three courses employed the same supplemental digital learning support system (anonymized for review) integrated with each university’s LMS, designed to support students’ monitoring of their learning through structured opportunities for deliberate practice, feedback, and self-reflection aligned with course objectives.


Students had access to three types of practice opportunities: (i) incentivized digital exercises, (ii) non-incentivized digital exercises, and (iii) incentivized paper-based exercises that were not integrated into the digital system.

Both the digital system and the paper-based exercises required students to apply their knowledge to simple and complex tasks, including transfer to new contexts; however, the formats differed in focus.
Digital exercises—some of which were also available on paper—were limited to tasks with clearly defined correct answers and therefore emphasized application or near transfer, whereas paper-based exercises more often required far transfer and more elaborate solution explanations.

Students could attempt digital exercises an unlimited number of times and received automatically generated response feedback immediately after every attempt. Paper-based exercises, by contrast, were graded by teaching assistants, resulting in delayed feedback.


Digital exercises were released shortly after almost every lecture and were either incentivized or non-incentivized, while paper-based exercises were released weekly (TCS-A) or biweekly (Logic-A and Logic-B).
To qualify for incentives, students were required to submit recent digital and paper-based exercises by the release date of the next paper-based assignment, resulting in one-week (TCS-A) or two-week (Logic-A and Logic-B) completion windows for paper-based exercises. Students were expected to submit solutions to digital exercises individually, but they could submit solutions to paper-based exercises in teams of up to three students.

Incentive structures differed across institutions. At University A (Courses TCS-A and Logic-A), students received bonus points toward their final exam grade for completing incentivized exercises. In contrast, at University B (Course Logic-B), students were required to accumulate a minimum number of points from incentivized tasks to become eligible for the final exam. See Table–\ref{tab:coursedifferences} for an overview.

\begin{table}
    \caption{Overview over the differences of the three studied courses.}
    \label{tab:coursedifferences}
    \newcolumntype{R}{>{\raggedright\arraybackslash}p{3.3cm}}
    \centering
    \footnotesize
    \begin{tabular}{lRRR}
        \toprule
        \multicolumn{4}{c}{\textbf{Course structure}}\\
        \midrule
         \textbf{Aspect} & \textbf{TCS-A} & \textbf{Logic-A} & \textbf{Logic-B}  \\
         \midrule
         Lecture & 2 x 90 min per week & 1 x 90 min per week & 1 x 90 min per week \\
         Exercise sessions & 90 min every week & 90 min every other week & 90 min every other week \\
         Tutorials & no & no & 90 min every week \\
         Slides available & yes & yes & yes \\
         Forum available & yes& yes & yes \\
         Assign. Incentivization & bonus points for exam & bonus points for exam & admission for exam \\
         \midrule
         \multicolumn{4}{c}{\textbf{Data usage}}\\
        \midrule
         \textbf{Aspect} & \textbf{TCS-A} & \textbf{Logic-A} & \textbf{Logic-B}  \\
         \midrule
         Clicks on slides available & yes & yes & no \\
         Clicks on forum available & yes & yes & no \\
         \bottomrule
    \end{tabular}
\end{table}

\subsection{Data Collection and Measures}


Behavioral trace data were extracted from two digital learning environments—the universities’ learning management system (Moodle) and the supplemental teaching support system [anonymized for blinded review]---which automatically logged students’ interactions (e.g., slide and forum clicks, submissions of digital and paper-based exercises) along with timestamps, providing a fine-grained record of when, how often, and in what sequence students engaged with course materials. 


Variable construction was guided by SRL theory and prior empirical work \parencite[e.g.,][]{zimmerman_becoming_2002, DuEA2023} theory-guided behavioral indicators from LMS trace data that can be interpreted in relation to self-regulated learning processes. Digital trace data record observable learning actions (e.g., accessing resources, submitting tasks, or interacting with course materials) and therefore provide behavioral evidence of learners’ engagement with learning tasks. However, such traces do not directly capture the underlying cognitive and metacognitive processes of SRL and must be interpreted through theoretical frameworks that link observable actions to regulatory activity \parencite[e.g.,][]{conijn_predicting_2017}.

Consistent with established SRL frameworks, which conceptualize learning as a cyclical process involving planning, monitoring, and regulation of effort and strategies, variables were organized into descriptive categories aligning with task engagement and coverage, temporal regulation (time management), post-completion engagement with practice tasks, engagement with course resources, engagement with help-seeking resources, and performance. 

Following recent work in learning analytics that emphasizes theory-aligned and context-responsive interpretation of digital traces, these categories represent observable behavioral indicators that can be mapped to SRL processes, while acknowledging that the underlying regulatory processes themselves remain only indirectly observable in LMS data \parencite[e.g.,][]{BernackiEA2025, Song.etal2025b, bernacki_clickstream_2026}. Table \ref{tab:variables} summarizes all derived variables within these SRL-related feature families, enabling theory-informed interpretation of digital traces. 

\paragraph{Engagement with Practice Tasks.}  
In our context, behavioral indicators of students’ work on practice tasks capture how extensively and consistently they engage with assigned exercises. Specifically, we used the proportion of submitted tasks across digital and paper-based exercises and the time spent on course pages as low-inference indicators of coverage and time investment that may afford monitoring and regulation but do not directly measure these metacognitive processes \parencite{DuEA2023}. Higher engagement levels, reflected in more frequent and sustained interactions, have been consistently associated with improved academic outcomes \parencite{VirvouEA2019, FinchamEA2019, vonKeyserlingk.etal2025}.

\textbf{Time Management.}  
Time management indicators describe the temporal regulation of learning activities, a central component of SRL associated with academic success \parencite{WoltersB2021}. We included variables such as time remaining before deadlines, timing of first engagement, number of weekdays spent on coursework, and weekly study sessions. Early task initiation and consistent engagement across weeks are generally predictive of stronger performance, whereas frequent last-day submissions are often interpreted as consistent with procrastination or poor regulation of effort \parencite{LiBW2020, MontgomeryEA2019, vonKeyserlingkEA2023, ArizmendiEA2023}. Early task initiation and consistent engagement across weeks may provide observable traces that index more effective temporal regulation.

\textbf{Post-completion review of practice tasks.}  
Post-completion review indicators capture students’ re-engagement with tasks after initial success, reflecting opportunities for monitoring and consolidation. These variables capture whether students re-engage with exercises or pages after a successful submission, for example by revisiting previously solved tasks or related materials. Because these events occur in spaces that provide feedback and worked solutions, they can afford opportunities for metacognitive processes such as monitoring and review.
Such behaviors may reflect adaptive regulation and goal-oriented persistence, both associated with deeper comprehension and long-term retention \parencite{FanEA2021, KizilcecEA2017, LanEA2019, MatchaEA2020, vanHalemEA2020}. However, any given re-engagement may also reflect other goals, like checking requirements.

\textbf{Lecture Slides and Forum Clicks.} 
Resource-access indicators reflect how students make use of instructional materials and peer support tools. Students in all courses had access to lecture slides and discussion forums within the LMS. Lecture slides were provided as instructor-generated materials designed to afford preview and review of lesson content, while discussion forums were designed to afford both organizational and content-related help-seeking and peer-instruction interaction.
However, in Course Logic-B, lecture slides and the forum were hosted externally, preventing the collection of click data. Instructors of the respective courses at Institution~A reported that the forums were typically used by only a few students and rarely for content-related questions.

\textbf{Performance Indicators.}  
Performance indicators were included to align behavioral traces with demonstrated learning outcomes. In addition to behavioral features, performance variables were derived to capture the quality of students’ work. Because students could attempt most digital exercises repeatedly until success and most students solved the tasks eventually, we used only performance scores from paper-based assignments, which were evaluated by teaching assistants. These indicators provide a direct measure of learning outcomes during the semester and enhance the interpretability of engagement patterns in relation to achievement \parencite{DuEA2023, SchwerterDBM2022}.

The variable families described above can be related to a targeted subset of SRL components across two SRL frameworks. Indicators of early task initiation---time to deadline and any interaction before the last day---can be related to temporal planning and forethought: in Zimmerman's (\citeyear{zimmerman_attaining_2000}) SRL model, these behaviors may reflect the goal-setting and strategic-planning processes of the forethought phase, while in Winne and Hadwin's (\citeyear{winne_studying_1998}) COPES model they correspond to Stage 2, in which students allocate effort and construct a plan under time constraints. Submission frequency and session regularity can be related to behavioral effort regulation and distributed engagement associated with the performance phase in Zimmerman's (\citeyear{zimmerman_attaining_2000}) model and with Stage 3 (enacting study tactics) in Winne and Hadwin's (\citeyear{winne_studying_1998}) model. Paper-based performance scores offer a more direct indicator of demonstrated learning outcomes and can be related to the self-evaluation component of Zimmerman's (\citeyear{zimmerman_attaining_2000}) self-reflection phase. We explicitly note that several SRL components---including metacognitive monitoring accuracy, goal-setting specificity, and self-motivational beliefs---remain only weakly or indirectly observable in LMS log data and cannot be inferred reliably from these indicators alone \parencite{BernackiEA2025, DuEA2023}.

\textbf{Variable Screening and Exclusion.} 
To ensure parsimony and model stability, highly correlated variables ($r > .90$) were inspected, and one variable from each correlated pair was removed to mitigate multicollinearity \parencite{lazar2012survey}. Excluded variables included: (i) proportion of tasks (not pages) submitted, (ii) indicator of all submissions completed before the last day, (iii) time on page after successful submission (minutes), (iv) proportion of correctly completed pages, (v) duplicate proportion of correctly completed pages, and (vi) proportion of correctly completed tasks on started pages.

\begin{table}[htpb]
\centering
\footnotesize
\caption{Overview of variables and explanations}
\label{tab:variables}

\begin{tabularx}{\textwidth}{p{6cm} l s X s} 
\toprule
\textbf{Variable} & \textbf{Abbreviation} & \textbf{Variable Type} &
\textbf{Explanation} & \textbf{References} \\
\midrule

\multicolumn{5}{l}{\textbf{Task Engagement and Coverage}} \\
\cmidrule{1-1}
Number of solution deliveries 
  & eng1 
  & 
  & Counts of solution submissions for digital practice tasks (relative to the cohort median).
  & \parencite{KimEA2019, LiBW2020, MontgomeryEA2019, VanAlten2020} \\

Time on page in minutes over all pages
  & eng5 
  & 
  & Total time spent on pages in the digital practice system, used as an indicator of behavioral engagement with course materials.
  & \parencite{KimEA2019, LiBW2020, MontgomeryEA2019, VanAlten2020, RodriguesEA2019, KizilcecEA2017} \\
Proportion of pages with submission 
  & par2 
  & 
  & Fraction of full pages for which a student submits a solution (coverage of assigned practice).
  & \parencite{RodriguesEA2019, KizilcecEA2017} \\
  
\multicolumn{5}{l}{\textbf{Time Management (Temporal Regulation and Pacing)}} \\
\cmidrule{1-1}
Time to deadline in days 
  & eng2 
  & 
  & Time between students’ first interaction with a task and its deadline, indexing early versus late initiation.
  & \parencite{LiBW2020} \\

Submission on the last day
  & eng3 
  & Indicator 
  & Indicates whether task submissions occurred on the deadline day.
  & \parencite{LiBW2020} \\

Any interaction before the last day
  & eng6 
  & Indicator
  & Indicates whether students engaged with a task before the final day.
  & \parencite{LiBW2020} \\

Number of days worked on tasks 
  & eng9 
  & 
  & Number of distinct days on which students engaged with tasks, reflecting behavioral regularity.
  & \parencite{MontgomeryEA2019, VanAlten2020, LeeR2017, RodriguesEA2019, LanEA2019} \\

Number of sessions per week / fortnight 
  & eng10
  & 
  & Frequency of study sessions over time; a new session begins after at least 1.5 hours of inactivity.
  & \parencite{MontgomeryEA2019, VanAlten2020, LeeR2017, RodriguesEA2019, LanEA2019} \\

\multicolumn{5}{l}{\textbf{Post-Completion Engagement}} \\
\cmidrule{1-1}
Any interaction after successful submission
  & eng7 
  & Indicator
  & Re-engagement with an exercise page after achieving a correct submission.
  & \parencite{FanEA2021, KizilcecEA2017, LanEA2019, MatchaEA2020, vanHalemEA2020} \\

\multicolumn{5}{l}{\textbf{Lecture Slides and Forum Clicks}} \\
\cmidrule{1-1}
Number of downloads of lecture slides 
  & lecture\_clicks 
  & 
  & Slide downloads are interactions with instructor-provided materials designed to afford preview and review of lecture content.
  & \\

Number of clicks in forum 
  & forum 
  & 
  & Interactions with the course discussion forum, a resource that can support organizational or content-related help-seeking.
  & 
  \\
\bottomrule
\end{tabularx}
\vspace{0.5ex}
\begin{minipage}{\textwidth}
\footnotesize
\textit{Note.} The abbreviations are used in the importance plots Figures~\ref{fig:importance_en_all}, \ref{fig:importance_rf_all}, and \ref{fig:importance_xgb_all} due to limited space. See complete list of generated variables prior to omission for multicollinearity in Figure~\ref{tab_fulllist}
\end{minipage}
\end{table}

\begin{figure}[htbp]
    \centering
    \caption{Exam distributions}
    \label{fig:examdistribution}
    \includegraphics[width=0.325\linewidth, page=1, clip, trim = 0 0 0 40]{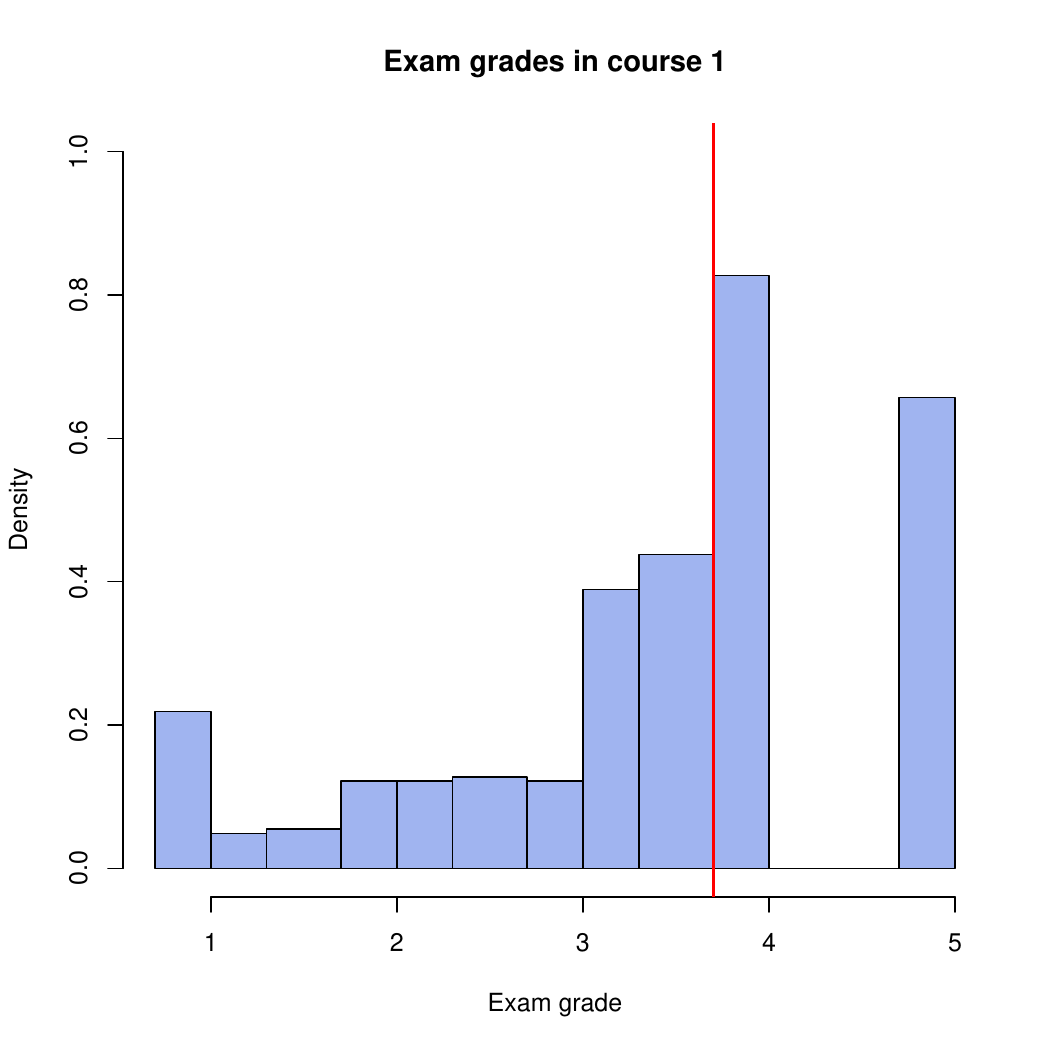}
    \includegraphics[width=0.325\linewidth, page=2, clip, trim = 0 0 0 40]{01_Figures/Klausurnoten.pdf} 
    \includegraphics[width=0.325\linewidth, page=3, clip, trim = 0 0 0 40]{01_Figures/Klausurnoten.pdf}
     \begin{flushleft}
\footnotesize \textit{Note.} Histograms display the distribution of final exam grades for each course. The dashed vertical line indicates the threshold used to classify students as “at risk” (grade $\geq$ 3.7). Grades follow the German grading scale ranging from 0.7 (best) to 5.0 (fail).
\end{flushleft}
\end{figure}

\textbf{Outcome variable}. The primary dependent variable was a binary indicator of academic risk, derived from students’ final exam grades. Exams were graded according to the German 5-point scale, ranging from 0.7 (best) to 5.0 (fail). Students receiving a grade of 3.7 or worse were classified as ``at-risk,'' as these grades represent either failure or marginal passing performance near the failure threshold. Figure~\ref{fig:examdistribution} displays the exam grade distributions across the three courses.

\subsubsection{Data aggregation}

To capture the temporal dynamics of students’ learning behavior, all indicators were computed on a weekly basis, distinguishing between task type (digital vs.\ paper-based), incentive condition (incentivized vs.\ non-incentivized), and timing relative to incentive deadlines (before vs.\ one week after; see Figure~\ref{fig:aggregation}). Exercises were released one week before their deadlines in Course TCS-A and two weeks before in Courses Logic-A and Logic-B. Only minimal activity occurred after deadlines, indicating that students rarely revisited earlier materials within the digital teaching support system.

Two complementary aggregation strategies were used to create cumulative predictors reflecting SRL-related patterns over time. The first strategy---progressive aggregation---expanded the observation window sequentially (e.g., Week 1; Weeks 1–2; Weeks 1–3; …; Weeks 1–12) and computed cumulative means and standard deviations to represent the evolving intensity and stability of behavioral engagement (see Figure~\ref{fig:aggregation}). This approach was chosen for three methodological reasons. First, progressive aggregation maintains a constant number of predictors across prediction weeks, which enables direct comparison of model performance as additional behavioral data become available. Second, aggregating behaviors cumulatively reduces interpretational complexity relative to models that include week-specific predictors, where the meaning of a variable may shift depending on the week in which it occurs. Third, cumulative indicators align conceptually with self-regulated learning frameworks that emphasize sustained patterns of engagement, effort regulation, and time management over isolated weekly events \parencite{DuEA2023, MatchaEA2020}. Consequently, progressive aggregation captures the accumulation of learning behaviors that may signal emerging regulatory strategies over the course of the semester. Compared with alternative approaches that rely on week-specific predictors or lagged models, this design reduces the risk that predictions are driven by idiosyncratic week-level fluctuations---such as a single week of elevated activity---that may not generalize across courses or institutions. By summarizing engagement as a running cumulative mean and standard deviation, the approach captures whether a student is, in general, engaging consistently and early, rather than whether they happened to do so in any particular week.

The second strategy---early-reset aggregation---introduced a temporal reset after the initial four weeks. In contrast to progressive aggregation, which accumulates information continuously over the semester, the early-reset strategy explicitly separates early-semester engagement from later behavioral patterns. Data were first aggregated for Weeks 1–4, followed by a new accumulation window beginning in Week 5. This design was motivated by empirical literature, which posits that the early term constitutes a critical initial behavioral phase during which regulatory behaviors are most malleable \parencite{BernackiEA2020, ArizmendiEA2023}. Consequently, Weeks 1–4 were ``frozen'' as a separate cumulative predictor set, capturing initial planning and early adaptation to course demands, while the reset window (Weeks 5–12) represented subsequent monitoring and control behaviors.

This design served two methodological purposes:
(i) to allow for varying effects of engagement and performance at the beginning compared with the rest of the semester, given that some studies might attempt to compensate for early procrastination; and
(ii) to reflect that the first four weeks typically introduce foundational concepts critical for later learning and performance.

For the early-reset strategy, performance-related variables were computed only for Weeks 1–4, as later performance becomes less actionable for formative intervention.

\begin{figure}[htbp]
    \centering
    \caption{Data aggregation strategies.}
    \label{fig:aggregation}

    \begin{subfigure}{\textwidth}
    \centering
    \caption{Progressive aggregation}
    \begin{tikzpicture}[font=\normalsize, node distance=1mm, scale=0.62, transform shape]
    \tikzset{
      box/.style={draw=utred, rounded corners=5pt,
                  line width=0.8pt, fill=white, inner sep=8pt,
                  minimum width=3cm, align=center},
      box_week/.style={draw=utred, rounded corners=5pt,
                  line width=0.8pt, fill=white, inner sep=4pt,
                  minimum width=6cm, align=center},
      box_aggre/.style={draw=utred, rounded corners=5pt,
                  line width=0.8pt, fill=white, inner sep=4pt,
                  minimum width=5cm, align=center},
      arrow/.style={->, >=latex, thick, draw=utblue}
    }

    \node[box] (prop) {\textbf{Proportion of tasks}};
    \node[box, above=2.0mm of prop, xshift=40mm] (inc) {Clicks of\\incentivized\\exercises};
    \node[box, below=0.5mm of prop, xshift=40mm] (non) {Clicks of\\non-incentivized\\exercises};

    \node[box_week, right=15mm of inc, yshift=15mm]   (w1i) {Exercises of week 1 in week 1};
    \node[box_week, right=15mm of inc, yshift=7.5mm]  (w2i) {Exercises of week 2 in week 2};
    \node[box_week, right=15mm of inc, yshift=0mm]    (w3i) {Exercises of week 3 in week 3};
    \node[right=30mm of inc, yshift=-7mm]             (dotsi) {\textbf{$\vdots$}};
    \node[box_week, right=15mm of inc, yshift=-15mm]  (w12i) {Exercises of week 12 in week 12};

    \node[box_week, right=15mm of non, yshift=7mm]    (w1n) {Exercises of week 1 in week 1};
    \node[right=30mm of non, yshift=1mm]              (dotsn) {\textbf{$\vdots$}};
    \node[box_week, right=15mm of non, yshift=-8mm]   (w12n) {Exercises of week 12 in week 12};

    \node[box_aggre, right=15mm of w1i]  (o1i)  {Week 1};
    \node[box_aggre, right=15mm of w2i]  (o2i)  {M \& SD of weeks 1--2};
    \node[box_aggre, right=15mm of w3i]  (o3i)  {M \& SD of weeks 1--3};
    \node[right=75mm of dotsi]           (odotsi) {\textbf{$\vdots$}};
    \node[box_aggre, right=15mm of w12i] (o12i) {M \& SD of weeks 1--12};

    \node[box_aggre, right=15mm of w1n]  (o1n) {Week 1};
    \node[right=75mm of dotsn]           (odotsn) {\textbf{$\vdots$}};
    \node[box_aggre, right=15mm of w12n] (o12n) {M \& SD of weeks 1--12};

    \draw[arrow] (prop.north) -- (inc.west);
    \draw[arrow] (prop.south) -- (non.west);

    \draw[arrow] (inc.east) -- (w1i.west);
    \draw[arrow] (inc.east) -- (w2i.west);
    \draw[arrow] (inc.east) -- (w3i.west);
    \draw[arrow] (inc.east) -- (w12i.west);

    \draw[arrow] (non.east) -- (w1n.west);
    \draw[arrow] (non.east) -- (w12n.west);

    \draw[arrow] (w1i.east) -- (o1i.west);
    \draw[arrow] (w1i.east) -- (o2i.west);
    \draw[arrow] (w1i.east) -- (o3i.west);
    \draw[arrow] (w1i.east) -- (o12i.west);

    \draw[arrow] (w2i.east) -- (o2i.west);
    \draw[arrow] (w2i.east) -- (o3i.west);
    \draw[arrow] (w2i.east) -- (o12i.west);

    \draw[arrow] (w3i.east) -- (o3i.west);
    \draw[arrow] (w3i.east) -- (o12i.west);
    \draw[arrow] (w12i.east)-- (o12i.west);

    \draw[arrow] (w1n.east) -- (o1n.west);
    \draw[arrow] (w1n.east)-- (o12n.west);
    \draw[arrow] (w12n.east)-- (o12n.west);
    \end{tikzpicture}
    \end{subfigure}

    \vspace{1em}

    \begin{subfigure}{\textwidth}
    \centering
    \caption{Early-reset aggregation}
    \begin{tikzpicture}[font=\normalsize, node distance=1mm, scale=0.62, transform shape]
 \tikzset{
  box/.style={draw=utred, rounded corners=5pt,
              line width=0.8pt, fill=white, inner sep=8pt,
              minimum width=3cm, align=center},
  box_week/.style={draw=utred, rounded corners=5pt,
              line width=0.8pt, fill=white, inner sep=4pt,
              minimum width=6cm, align=center},
  box_aggre/.style={draw=utred, rounded corners=5pt,
              line width=0.8pt, fill=white, inner sep=4pt,
              minimum width=5cm, align=center},
  box_frozen/.style={draw=utred, rounded corners=5pt,
              line width=0.8pt, fill=gray!15, inner sep=4pt,
              minimum width=5cm, align=center},
  arrow/.style={->, >=latex, thick, draw=utblue},
  brace_label/.style={font=\small\itshape, align=center}
}

\node[box] (prop) {\textbf{Proportion of tasks}};
\node[box, above=20.0mm of prop, xshift=40mm] (inc) {Clicks of\\incentivized\\exercises};
\node[box, below=20.0mm of prop, xshift=40mm] (non) {Clicks of\\non-incentivized\\exercises};

\node[box_week, right=15mm of inc, yshift=22mm]   (w1i) {Exercises of week 1 in week 1};
\node[box_week, right=15mm of inc, yshift=13mm]   (w2i) {Exercises of week 2 in week 2};
\node[box_week, right=15mm of inc, yshift=4mm]    (w3i) {Exercises of week 3 in week 3};
\node[box_week, right=15mm of inc, yshift=-5.5mm]   (w4i) {Exercises of week 4 in week 4};
\node[right=70mm of inc, yshift=-14mm, font=\small\bfseries, text=gray] (divlabel) {\textbf{--- reset ---}};
\node[box_week, right=15mm of inc, yshift=-23mm]  (w5i) {Exercises of week 5 in week 5};
\node[box_week, right=15mm of inc, yshift=-32mm]  (w6i) {Exercises of week 6 in week 6};
\node[right=30mm of inc, yshift=-40mm]              (dotsn) {\textbf{$\vdots$}};
\node[right=110mm of inc, yshift=-40mm]              (dotsn) {\textbf{$\vdots$}};
\node[box_week, right=15mm of inc, yshift=-49mm]  (w12i) {Exercises of week 12 in week 12};

\node[box_week, right=15mm of non, yshift=10mm]   (w1n) {Exercises of week 1 in week 1};
\node[right=30mm of non, yshift=3mm]              (dotsn1) {\textbf{$\vdots$}};
\node[box_week, right=15mm of non, yshift=-6mm]   (w4n) {Exercises of week 4 in week 4};
\node[right=70mm of non, yshift=-12mm, font=\small\bfseries, text=gray] (divlabeln) {--- reset ---};
\node[box_week, right=15mm of non, yshift=-20mm]  (w5n) {Exercises of week 5 in week 5};
\node[right=30mm of non, yshift=-27mm]            (dotsn2) {\textbf{$\vdots$}};
\node[box_week, right=15mm of non, yshift=-34mm]  (w12n) {Exercises of week 12 in week 12};

\node[box_frozen, right=15mm of w1i]  (o1i)  {Week 1};
\node[box_frozen, right=15mm of w2i]  (o2i)  {M \& SD of weeks 1--2};
\node[box_frozen, right=15mm of w3i]  (o3i)  {M \& SD of weeks 1--3};
\node[box_frozen, right=15mm of w4i]  (o4i)  {\shortstack{M \& SD of weeks 1--4\\[0pt](frozen)}};
\node[box_aggre, right=15mm of w5i]  (o5i)   {Week 5};
\node[box_aggre, right=15mm of w6i]  (o6i)   {M \& SD of weeks 5--6};
\node[box_aggre, right=15mm of w12i]  (o12i)   {M \& SD of weeks 5--12};

\node[box_frozen, right=15mm of w1n]  (o1n)  {Week 1};
\node[right=75mm of dotsn1]           (odotsn1) {\textbf{$\vdots$}};
\node[box_frozen, right=15mm of w4n]  (o4n)  {\shortstack{M \& SD of weeks 1--4\\(frozen)}};
\node[box_aggre,  right=15mm of w5n]  (o5n)  {Week 5};
\node[right=75mm of dotsn2]           (odotsn2) {\textbf{$\vdots$}};
\node[box_aggre,  right=15mm of w12n]  (o12n)  {Week 12};

\draw[arrow] (prop.north) -- (inc.west);
\draw[arrow] (prop.south) -- (non.west);

\draw[arrow] (inc.east) -- (w1i.west);
\draw[arrow] (inc.east) -- (w2i.west);
\draw[arrow] (inc.east) -- (w3i.west);
\draw[arrow] (inc.east) -- (w4i.west);
\draw[arrow] (inc.east) -- (w5i.west);
\draw[arrow] (inc.east) -- (w6i.west);
\draw[arrow] (inc.east) -- (w12i.west);

\draw[arrow] (non.east) -- (w1n.west);
\draw[arrow] (non.east) -- (w4n.west);
\draw[arrow] (non.east) -- (w5n.west);
\draw[arrow] (non.east) -- (w12n.west);

\draw[arrow] (w1i.east) -- (o1i.west);
\draw[arrow] (w1i.east) -- (o2i.west);
\draw[arrow] (w1i.east) -- (o3i.west);
\draw[arrow] (w1i.east) -- (o4i.west);
\draw[arrow] (w2i.east) -- (o2i.west);
\draw[arrow] (w2i.east) -- (o3i.west);
\draw[arrow] (w2i.east) -- (o4i.west);
\draw[arrow] (w3i.east) -- (o3i.west);
\draw[arrow] (w3i.east) -- (o4i.west);
\draw[arrow] (w4i.east) -- (o4i.west);

\draw[arrow] (w5i.east) -- (o5i.west);
\draw[arrow] (w5i.east) -- (o6i.west);
\draw[arrow] (w5i.east) -- (o12i.west);
\draw[arrow] (w6i.east) -- (o6i.west);
\draw[arrow] (w6i.east) -- (o12i.west);
\draw[arrow] (w12i.east) -- (o12i.west);

\draw[arrow] (w1n.east) -- (o1n.west);
\draw[arrow] (w1n.east) -- (o4n.west);
\draw[arrow] (w4n.east) -- (o4n.west);
\draw[arrow] (w5n.east) -- (o5n.west);
\draw[arrow] (w5n.east) -- (o12n.west);
\draw[arrow] (w12n.east) -- (o12n.west);

\draw [decorate, decoration={brace, amplitude=6pt, raise=3pt},
       draw=gray!70]
  (o1i.north east) -- (o4i.south east)
  node[midway, right=10pt, brace_label] {Early-semester\\summary\\(Weeks 1--4)};

\draw [decorate, decoration={brace, amplitude=6pt, raise=3pt},
       draw=gray!70]
  (o5i.north east) -- (o12i.south east)
  node[midway, right=10pt, brace_label] {Reset window\\(Weeks 5+)};

    \end{tikzpicture}
    \end{subfigure}
         \begin{flushleft}
\footnotesize \textit{Note.} Panel (a) illustrates progressive aggregation, in which behavioral indicators are accumulated sequentially from the beginning of the semester (e.g., Week 1; Weeks 1–2; Weeks 1–3). Panel (b) illustrates early-reset aggregation, in which Weeks 1–4 are first aggregated into a fixed early-semester summary and a new cumulative window begins in Week 5. For visual clarity, panel (b) illustrates the reset logic through Week 6.
\end{flushleft}
\end{figure}



\subsection{Analytic approach}

We implemented a predictive modeling framework to identify students (at risk of) failing the final exam. Three binary classification algorithms were selected to balance interpretability and predictive power: Elastic Net logistic regression \parencite{friedman_elements_2001}, Random Forest in its Probability Forest variant \parencite{malley2012probability}, and Extreme Gradient Boosting (XGBoost) \parencite{ChenG2016}. This combination allowed for direct comparison between a regularized linear model and two nonlinear ensemble methods that can capture higher-order interactions among features.

We conducted both in-sample and out-of-sample validations, training and testing models within the same course and assessing generalizability across courses and institutions. Because the prevalence of at-risk students varied considerably (TCS-A: 62\%; Logic-A: 56 \; Logic-B: 14\%), two thresholding strategies were compared: (i) applying the classification threshold from the training course and (ii) recalibrating the threshold based on the target course’s at-risk proportion. The latter assumes perfect prior knowledge and represents an optimistic upper bound---analogous to knowing the failing-grade distribution at semester start---whereas the former provides a conservative generalization test.

We evaluated predictive performance using both threshold-independent and threshold-dependent measures, as no single metric fully captures how well early-warning models support instructional decisions. The area under the curve (AUC) reflects how well a model distinguishes higher- and lower-risk students overall, regardless of where a decision threshold is set. To evaluate performance at a specific decision point, we additionally report accuracy, which indicates the proportion of students correctly classified overall; sensitivity (recall), which reflects how well the model identifies students who are actually at risk; and specificity, which reflects how well the model avoids incorrectly flagging students who are not at risk. Because accuracy alone can be misleading when risk prevalence is uneven, we also report the $F_1$ score, which summarizes the balance between identifying at-risk students and avoiding false alarms \parencite{powers2011evaluation}, and Cohen’s~$\kappa$, which reflects agreement between predicted and observed outcomes beyond chance \parencite{ArizmendiEA2023}.

In terms of interpretation for instructional use, these complementary measures capture different risks associated with acting on predictions. AUC indicates whether a model can reliably rank students by risk in general. Accuracy summarizes overall correctness but can mask important errors when at-risk students are rare or common. Sensitivity and specificity make these trade-offs explicit by showing how well a model identifies students who need support while avoiding unnecessary intervention for those who do not. The $F_1$ score summarizes how well the model balances identifying students who truly need support against over-identifying students who do not, focusing attention on performance where instructional decisions are most consequential. Cohen’s~$\kappa$ provides an additional safeguard by indicating whether apparent accuracy reflects meaningful prediction rather than agreement driven by chance or base-rate effects. Considered jointly, these measures allow instructors to judge not only whether a model predicts well, but whether it does so in a way that supports timely, fair, and efficient allocation of limited support resources.

Feature importance was assessed via standardized coefficients (Elastic Net), Gini impurity (RF), and out-of-bag permutation importance (XGB) within each week. 

To enhance interpretability for intervention design, predicted probabilities were recalibrated using Platt scaling \parencite{PlattEA1999}. Calibration corrects systematic biases in probability estimates and ensures that predicted risk levels correspond to observed failure rates---critical for allocating limited support resources equitably \parencite{PhelpsEA2024}. Calibration quality was evaluated both in-sample and out-of-sample. See Supplementary Online Materials for more technical details of the analytical approach.

All analyses were performed in R 4.4.3 \parencite{RCoreTeam2025} using the \texttt{mlr3} ecosystem \parencite{mlr3, mlr3learners, mlr3tuning, mlr3filters}. Specific learners were implemented via \texttt{glmnet} for Elastic Net \parencite{FriedmanHT2010}, \texttt{ranger} for Random Forest \parencite{WrightZ2017}, and \texttt{xgboost} for gradient boosting \parencite{ChenEA2024}. Data preprocessing and evaluation used \texttt{tidyverse} \parencite{tidyverse}, \texttt{caret} \parencite{Kuhn2008}, and \texttt{pROC} \parencite{pROC}, while probability calibration and visualization were conducted with \texttt{platt} \parencite{platt} and \texttt{CalibrationCurves} \parencite{DeCock_CalibrationCurves_2023}.

\section{Results}

\subsection{Student Participation}

\begin{figure}[htb]
\centering
\caption{Weekly submission rates.} \label{fig:submission_rates}
\includegraphics[width=0.99\textwidth]{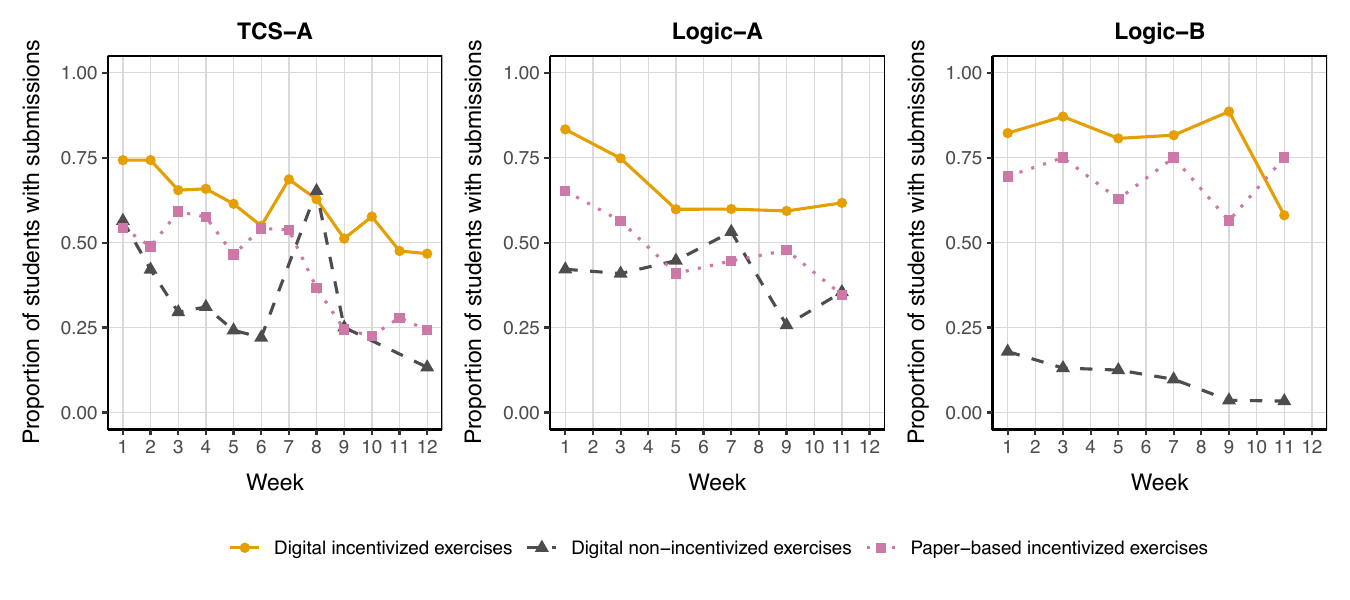} \\
\vspace{-1.5em}
\begin{flushleft}
\footnotesize{\textit{Note.} Lines represent the proportion of students submitting each type of assignment in a given week. Digital incentivized exercises provide bonus points or admission eligibility for the final exam, whereas digital non-incentivized exercises are optional practice tasks. Paper-based exercises are graded assignments submitted offline. In Logic-A and Logic-B, assignments were scheduled only for odd-numbered weeks.}
\end{flushleft}
\end{figure}

Across all courses, students completed most incentivized assignments, though submission rates declined gradually over the semester (see Figure~\ref{fig:submission_rates}). Completion rates for incentivized digital and paper-based exercises were comparable across contexts, whereas participation with non-incentivized digital tasks showed more pronounced variation. The two courses at University A (TCS-A and Logic-A) displayed consistently moderate participation in non-incentivized tasks across the term, with roughly one-quarter to one-half of students completing these exercises each week. In contrast, students at University B (Logic-B) engaged only minimally with non-incentivized tasks, with participation rarely exceeding 10–18\%. Despite these differences in optional-task participation, incentivized assignments elicited high and relatively stable completion rates in all three courses. At University B, students were required to accumulate a minimum number of points from incentivized tasks to qualify for the exam. Consistent with this policy, the instructor reported that students were more likely to stop submitting incentivized tasks once they had secured the necessary points. The late-term decline in submissions aligned with the point at which most students had already satisfied the exam-admission requirement.

\label{sec:results}

\subsection{How well do SRL-aligned predictive models generalize within and across courses and institutions? (RQ1)}

\subsubsection{Predictive Model Performance Over Time (RQ1a)}

\begin{figure}[htbp]
  \centering
  \caption{Weekly model quality metrics for Elastic Net, XGBoost, and Random Forest --- Course Logic-A as reference}
  \label{fig:gesamt_ref_course2_os1_course1_os2_course3}
  \includegraphics[width=0.99\linewidth]{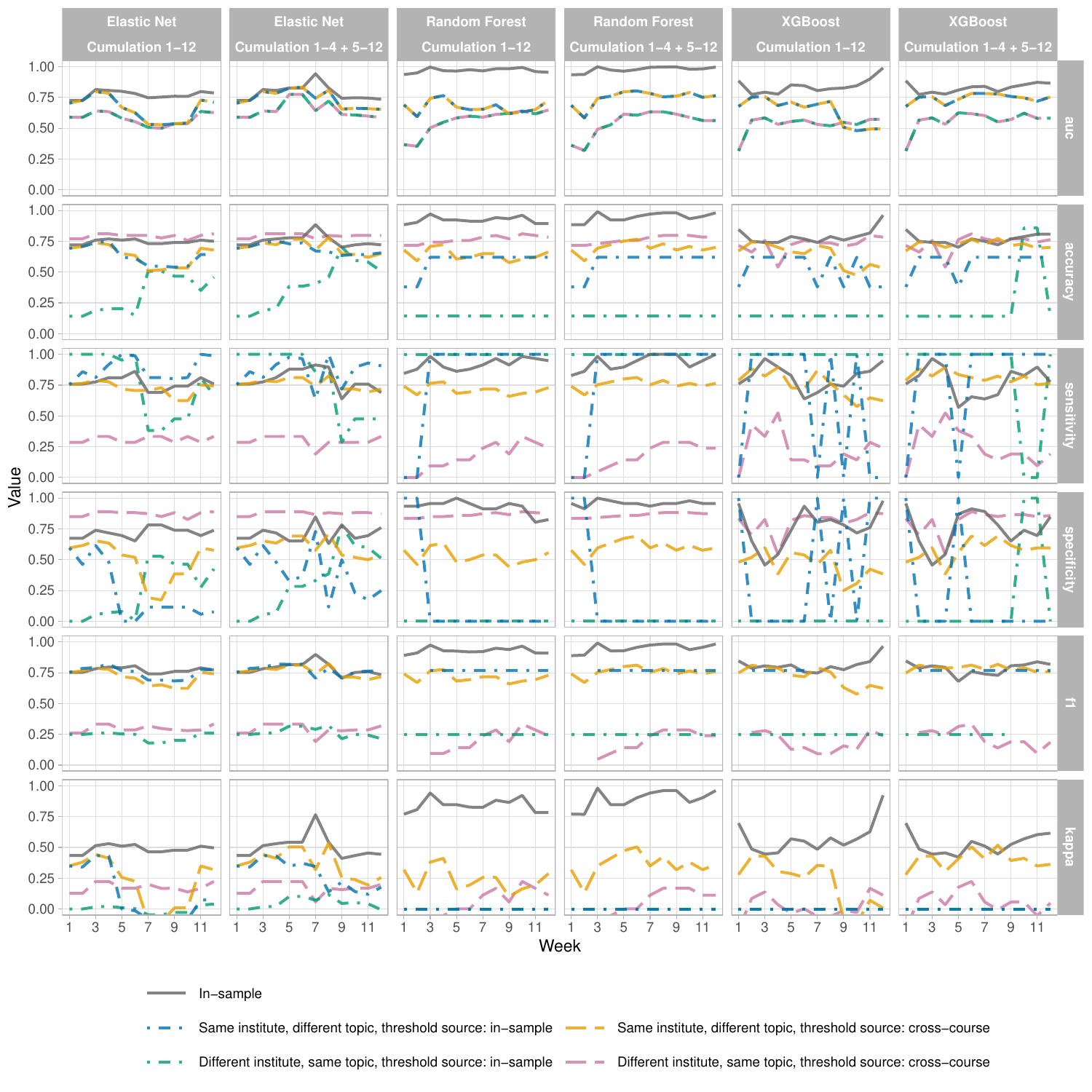}
  \begin{flushleft}
\footnotesize{\textit{Note.} Columns represent predictive models (Elastic Net, Random Forest, and XGBoost) under two feature aggregation strategies: progressive aggregation and early-reset aggregation. Rows display evaluation metrics (AUC, accuracy, sensitivity, specificity, F1 score, and Cohen’s $\kappa$). The gray line indicates in-sample prediction performance for the training course. Colored lines represent out-of-sample predictions when models are transferred across courses or institutions. Line types distinguish threshold strategies: thresholds derived from the training course (source threshold) versus thresholds adjusted to match the at-risk prevalence of the target course.}
\end{flushleft}
\end{figure}

We first examined how predictive model performance evolved as additional weekly behavioral data became available. Figure~\ref{fig:gesamt_ref_course2_os1_course1_os2_course3} illustrates model performance using Course Logic-A as the reference with source-threshold transfer. The first, third, and fifth columns present results for Elastic Net, Random Forest, and XGBoost under progressive feature aggregation, while the second, fourth, and sixth column present results for Elastic Net, Random Forest, and XGBoost under the early-reset aggregation, respectively. Rows display the different evaluation metrics. Within each panel, the gray reference line represents in-sample predictions, and the colored lines depict out-of-sample predictions across courses. Because the proportion of at-risk students was measured at the end of the semester, the outcome variable remained constant across weeks. Thus, any changes in performance metrics over time reflect the additional predictive information gained from accumulating behavioral data rather than shifts in the underlying target distribution.

In-sample AUC values ranged between .73 and .81 for Elastic Net, between .94 and 1.00 for Random Forest, and between .77 and .99 for XGBoost. The steepest AUC increase for Elastic Net occurred within the first four weeks (from .73 to .81) and then stabilized between .75 and .80. Random Forest performance remained consistently high throughout the semester, whereas XGBoost showed a brief decline (from .89 to .77) mid-semester, followed by a strong recovery toward the end (up to .99). Despite these fluctuations, all models achieved high levels of discrimination, with Random Forest yielding the highest AUC values, followed by XGBoost and then Elastic Net.

A consistent trend also emerged across the additional evaluation metrics---accuracy, F$_1$, and Cohen's~$\kappa$. Accuracy and F$_1$ closely mirrored AUC trajectories. Cohen's~$\kappa$ displayed greater week-to-week variability, underscoring its sensitivity to small fluctuations in the relative balance of predicted positive and negative classifications as additional weekly information were added. Importantly, Cohen's~$\kappa$ values below zero were observed, reflecting cases where agreement fell below chance level; however, for visual consistency across models and courses, negative Cohen's~$\kappa$ estimates were not mapped in the graphs.

In contrast, the Random Forest and late-semester XGBoost models in our study achieved notably higher in-sample performance, with AUC values approaching .90, F$_1$ scores up to .97, and Cohen's~$\kappa$ values up to .94, exceeding benchmarks reported in prior literature.

Our results compare favorably with prior research employing digital trace data for predictive modeling. In their systematic review, \textcite{ArizmendiEA2023} reported that most studies achieved average prediction accuracies around 72\% (SD = 0.10), typically relying on accuracy alone as the main performance metric. Similarly, \textcite{BernackiEA2020} demonstrated that an early-warning model based on LMS engagement correctly identified approximately 75\% of students at risk of earning below a B in a large STEM lecture course. These comparisons should be interpreted with caution given contextual and methodological differences across studies. Nonetheless, the findings collectively underscore the robust predictive value of ensemble methods and highlight the importance of evaluating multiple complementary performance metrics beyond accuracy alone.

Further, using the second feature aggregation strategy in columns two, four, and six, we froze cumulative data from Weeks 1–4 and then reinitiated aggregation for Weeks 5–12 while retaining the early-term cumulative indicators as a stable baseline. Under this configuration, the relative model ranking remained largely consistent with the first strategy: Random Forest continued to perform at or near ceiling levels, Elastic Net maintained moderate accuracy but exhibited a distinct performance spike around Week 7, and XGBoost remained strong, albeit slightly below its late-term peak from aggregation strategy 1. Although this configuration preserved the overall pattern of model performance, several metrics---especially sensitivity and specificity---displayed pronounced volatility in Weeks 10–12 (see the late-term fluctuations in Figure~\ref{fig:gesamt_ref_course2_os1_course1_os2_course3}). This instability aligns with the markedly reduced amount of new behavioral data available at the end of the semester and the resulting amplification of class-imbalance effects, both of which can cause threshold-based metrics to swing sharply even when the underlying model structure remains unchanged. Overall, these late-term oscillations did not materially alter the broader pattern: dividing features into early-term and later-term segments introduced only marginal changes to predictive performance, and the primary predictive signal remained concentrated in the initial weeks of the semester.

\subsubsection{Generalizability Across Courses and Thresholding (RQ1b)}

Across all models, out-of-sample predictions were consistently lower than in-sample performance and varied depending on both the prediction week and the target course. 
Cross-course predictions were moderately accurate early and late in the semester for Elastic Net, but mid-semester generalization dropped to near-chance levels. Conversely, cross-university predictions (i.e., Course Logic-B), where base rates of at-risk students differed substantially, were generally close to random classification. 
Random Forest and XGBoost, despite their near-ceiling in-sample performance, did not translate into superior out-of-sample generalization. 
Figure 5 illustrates these patterns across prediction weeks, and a more detailed breakdown of model-specific trajectories is provided in the supplementary online materials.

The findings indicate that Elastic Net exhibited the most stable performance across courses and institutions, while the ensemble models’ strong in-sample accuracy failed to generalize, often collapsing to near-trivial classification when base-rate differences were pronounced.
Still, Elastic Net produced meaningful above-chance predictions early in the semester cross-course, the period most relevant for early intervention.

\paragraph{Changing classification threshold strategy.} Adjusting classification thresholds to reflect the target course’s base rate substantially improved out-of-sample prediction performance across courses, without altering the underlying discrimination of the models. As expected, AUC remained unchanged because threshold adjustment does not affect the ranking of predicted probabilities. However, operating-point performance improved markedly when thresholds were calibrated to the target context rather than transferred unchanged from the training course.

These results help explain the generalization patterns observed before. When models were applied to courses with substantially different at-risk prevalence---particularly in cross-university settings---fixed thresholds led to near-trivial classification. In contrast, base-rate–aware threshold calibration partially restored meaningful classification by aligning decision boundaries with the target population.
Together, these findings suggest that poor cross-context performance reflects not only limits of model transfer, but also the consequences of uninformed thresholding under shifting prevalence conditions.

\paragraph{Changing reference course.} Contrary to our initial expectation, using Course Logic-A as the reference which overlaps with Course TCS-A through its shared bachelor’s program and with Course Logic-B through its shared topic was not universally optimal for out-of-sample prediction (see Figures \ref{fig:gesamt_ref_course1_os1_course2_os2_course3} and \ref{fig:gesamt_ref_course3_os1_course1_os2_course2}). Changing the reference course influenced cross-course generalizability, but the resulting improvements were modest and method-dependent.

These findings reinforce that (i) the choice of reference course can slightly shift out-of-sample discrimination, particularly for Elastic Net, and (ii) the primary constraint on cross-course performance under threshold strategy 1 lies in the limited portability of thresholds rather than limited discriminative power. At the same time, the results also highlight a broader practical limitation: the three courses differ in workload structure, pacing, and engagement patterns, and these contextual factors shape the distribution of behavioral traces available for modeling. When selecting a reference course for reapplying prediction models, features of the learning task’s instructional design, topical focus, and broader institutional context (e.g., enrollment and participation structures) can therefore influence the quality of cross-course prediction. We elaborate on these dimensions of transfer in the Discussion. This contextual perspective also motivates the next set of results, where we examine which predictors---and which SRL processes---they reflect across courses.


\subsubsection{Key Predictors and SRL Processes (RQ1c–RQ1d)}

\begin{figure}[htbp]
  \centering
    \caption{Variable importance across weeks and courses -- Elastic Net all courses}
  \label{fig:importance_en_all}
\includegraphics[width=\linewidth, page = 4]{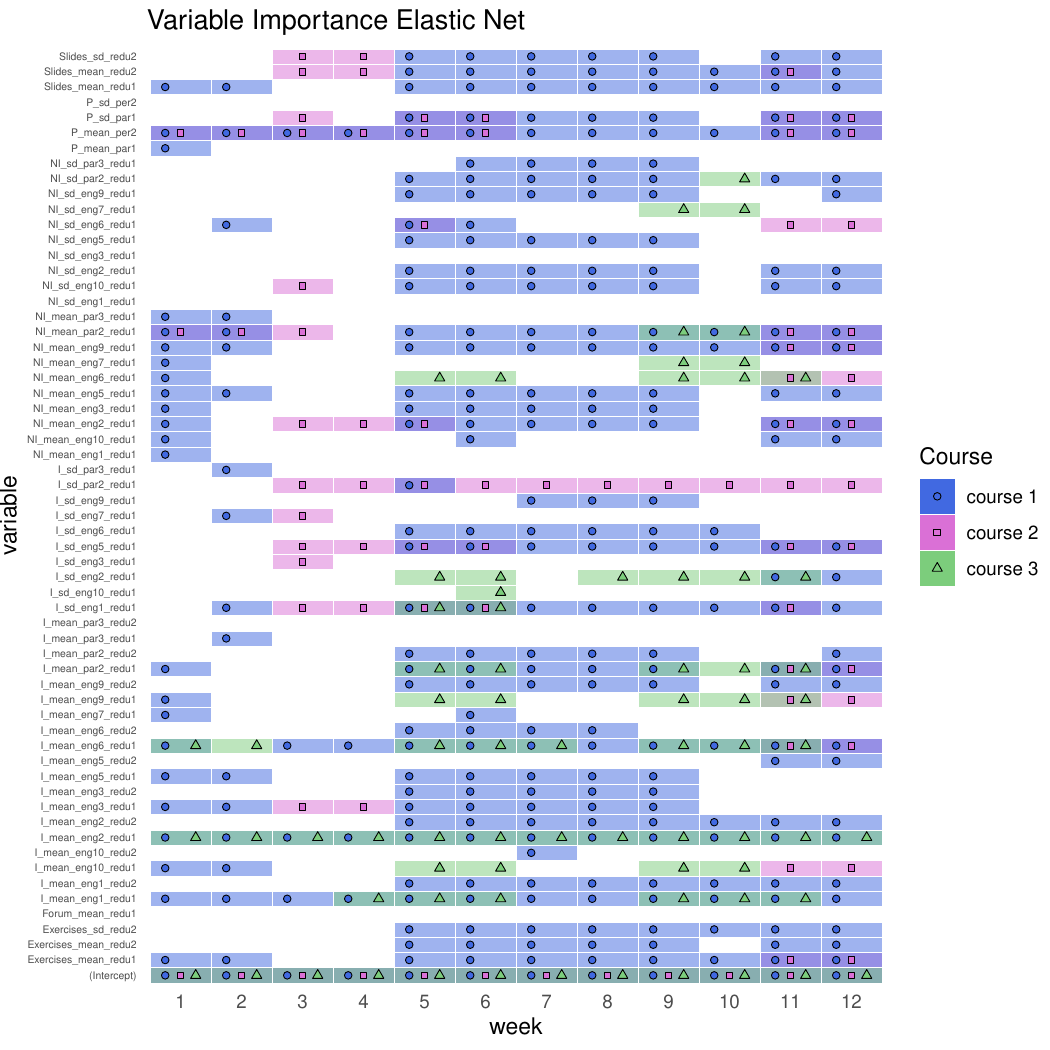}
\vspace{0.5ex}
\begin{minipage}{\textwidth}
\footnotesize
\textit{Note.} Incentivized problems start with ``I'', non-incentivized tasks start with ``NI'', and ``P'' denotes paper-based exercises. ``Mean'' and ``SD'' indicate whether aggregation is based on the mean or the standard deviation. ``redu1'' refers to tasks completed before the deadline, whereas ``redu2'' refers to tasks completed after the deadline. ``Exercises'' denotes the weighted points a student obtained in a given week for the paper-based exercises.
\end{minipage}
\end{figure}

Figure~\ref{fig:importance_en_all} presents the predictors selected by the Elastic Net models per week. Focusing first on Course Logic-A (purple entries), the model selected a relatively sparse set of predictors across weeks. The coefficients most consistently retained fall into three SRL-aligned families of variables:
(i) task and slide engagement within the upload week---that is, the mean number of clicks on exercises and lecture slides during the week in which materials were posted (the week-of-upload window, before the deadline);
(ii) engagement timing and frequency within the incentivized practice system, including the number of submissions, sessions per week or fortnight, average time to deadline, and whether students interacted with materials before the last day; and
(iii) participation and performance indicators, such as the proportion of pages with a submission and the proportion of tasks or pages completed correctly.


When comparing across courses, a clear pattern emerges in model sparsity: Course TCS-A displays substantially more variable selections across weeks, whereas Courses Logic-A and Logic-B show fewer weekly inclusions. Critically, week-specific overlap across all three courses is minimal. Applying the strict criterion of same variable, in the same week, across all courses, only three variables were jointly selected: (i) variability in the number of submissions (before the deadline, week of upload) during weeks 5--6; (ii) mean proportion of pages with a submission (before the deadline, week of upload) in week 11; and (iii) any interaction before the last day (mean, before the deadline, week of upload) also in week 11.

Outside these isolated overlaps, the exact timing and week-level selection of predictors diverged substantially across contexts. However, when collapsing across the semester---considering whether a predictive SRL-aligned indicator appeared at least once---the family-level consistency increased markedly. Across all three courses, the same broad SRL-related clusters recurred:
(i) steady engagement with core materials (tasks and slides accessed before deadlines); (ii) temporal regulation and frequency of engagement (number of submissions, sessions, time to deadline, early activity); and (iii) active participation and task performance (pages or tasks submitted and solved correctly).

For Random Forest, Figure~\ref{fig:importance_rf_all} corroborates the same underlying SRL-related feature families but displays a distinct importance distribution. Random Forest spreads predictive weight across a broader set of features, yielding a flatter importance profile. This difference is particularly evident when contrasting Course Logic-A and Course Logic-B. In the latter, participation and performance variables (pages submitted; tasks/pages correct) consistently rank among the most influential predictors across multiple weeks, suggesting a greater reliance on completion- and performance-based cues in that instructional context. 



For XGBoost (Figure~\ref{fig:importance_xgb_all}), two notable distinctions emerge: first, XGBoost concentrates importance more sharply within a small set of features each week, producing a more peaked profile of variable relevance. Second, because the model does not yield signed coefficients, it indicates how strongly a feature distinguishes students rather than whether higher values predict success or risk. 


Together, these findings indicate that although the timing of predictive signals varies by course, week, and method, the same underlying SRL processes---particularly time management, effort regulation, and sustained engagement---consistently drive prediction across contexts. Across models and courses, early and distributed activity, multiple study sessions, and submissions well before the deadline are the most reliable predictors of success. Similarly, consistent participation and performance, reflected in regular task completion and accuracy, contribute strongly to model performance. Conversely, deadline-concentrated or post-deadline behaviors are less informative and, when directional estimates are available (as in Elastic Net), are often associated with lower predicted achievement.

This convergence at the level of SRL-related feature families—despite substantial week-to-week variability—echoes complexity-oriented accounts of learning systems, in which regulatory patterns can remain stable even as their temporal expression shifts with contextual demands \parencite{hilpert_leveraging_2023}. Consistent with this view, interaction-dominant metrics in complexity approaches emphasize the structure and regularity of engagement rather than the specific tools used or the exact temporal sequence \parencite{hilpert_leveraging_2023}. 
Accordingly, SRL constructs such as time management, effort regulation, and sustained engagement may function as robust, transferable indicators of student success in digital learning environments.

\subsection{Does post-hoc probability calibration improve the accuracy of risk estimation for identifying students most in need of timely, targeted learning support? (RQ2)}

\begin{figure}[htbp]
  \centering
  \caption{Calibration plots (Week 3 \& 6, Elastic Net)}
  \label{fig:calibration_en}
    \begin{subfigure}{\linewidth}
    \centering
    \caption{Week 3}
  \includegraphics[width=\linewidth]{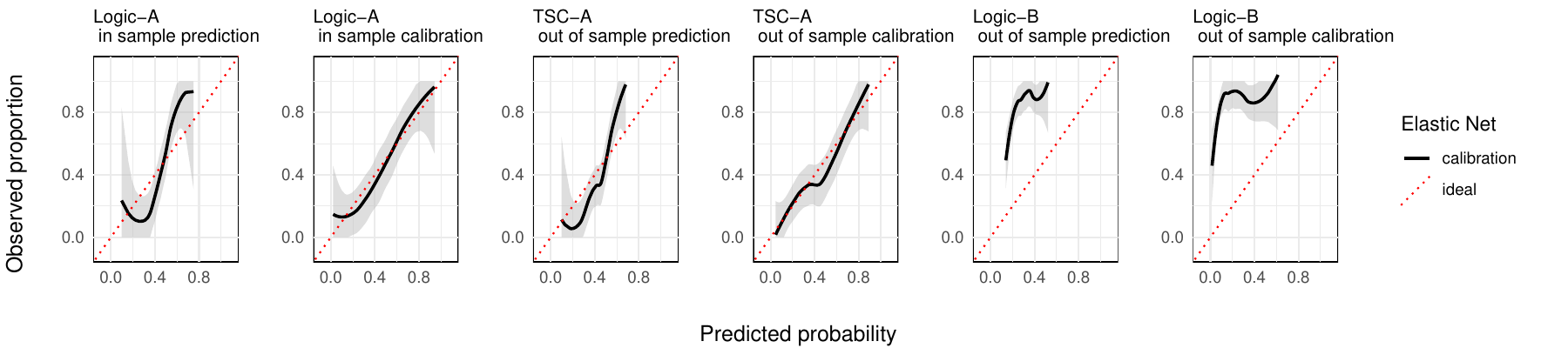}
    \end{subfigure}
  \begin{subfigure}{\linewidth}
  \centering
  \caption{Week 6}
  \includegraphics[width=\linewidth]{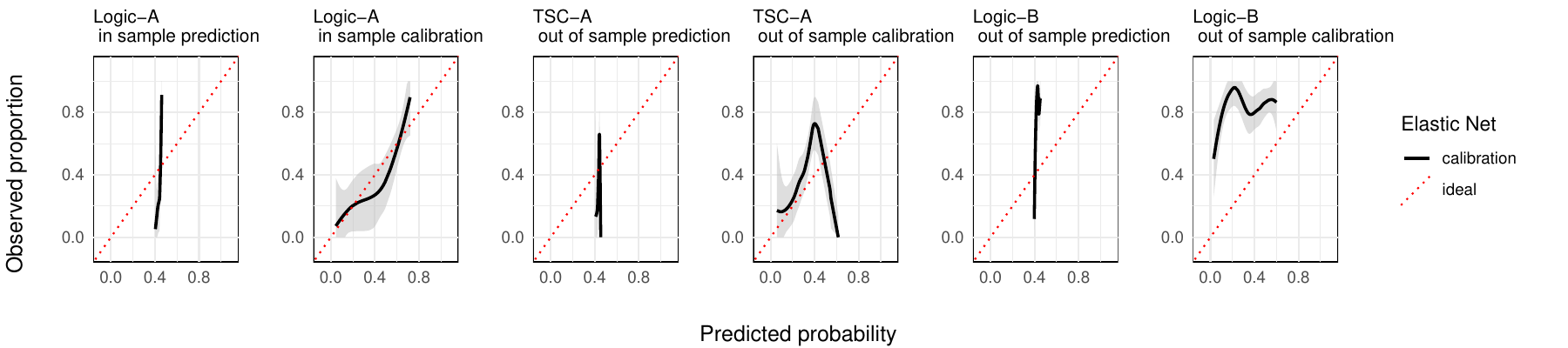}
  \end{subfigure}
  \begin{flushleft}
      \footnotesize \textit{Note}. Calibration curves compare predicted probabilities with observed outcome frequencies. The dashed diagonal line represents perfect calibration. Solid curves show empirical calibration before and after applying Platt scaling. Panels illustrate results for Week 3 and Week 6 predictions across in-sample and out-of-sample settings.
  \end{flushleft}
\end{figure}

Finally, we evaluated how accurately the models’ predicted probabilities reflected actual risk by applying Platt scaling to calibrate predicted outputs. Figures~\ref{fig:calibration_en} display calibration curves for Weeks 3 and 6 using Elastic Net as illustrative examples. Results for Random Forest, and XGBoost are shown in Supplementary Online Materials (Figures \ref{fig:calibration_rf}, and \ref{fig:calibration_xg}). Consistent with the discrimination results, in-sample predictions (Course Logic-A) already demonstrated high performance, though the uncalibrated probabilities tended to be either over- or under-confident. Applying probability calibration systematically improved alignment between predicted and observed outcome frequencies, shifting estimates closer to the diagonal reference line. The adjustment was most pronounced for Random Forest and XGBoost, which frequently produced overconfident risk estimates, whereas Elastic Net generated more moderate probabilities requiring smaller corrections, particularly in earlier weeks.

For out-of-sample prediction, a distinct contrast emerged between Course TCS-A and Course Logic-B. When transferring models within the same institution (Courses TCS-A and Logic-A), calibration notably improved probability accuracy, yielding curves that remained close to the diagonal even after cross-course transfer. In contrast, when transferring across institutions (Course Logic-A to Logic-B, University B), all models exhibited systematic miscalibration, consistently overestimating students’ risk levels. Even after applying Platt scaling, predicted probabilities for Course Logic-B remained inflated, reflecting the substantially lower proportion of at-risk students in that course.

A temporal pattern also emerged: calibration was more effective in earlier weeks ((and 4; see Supplementary Online Materials) of the semester, with curves closely approximating perfect calibration, whereas deviations widened in later weeks---particularly for out-of-sample predictions in Course Logic-B. 
One plausible interpretation, consistent with complexity-based perspectives, is that as students settle into course demands, engagement patterns become more regular, improving within-context discrimination while amplifying miscalibration when transferred across contexts with different base rates and participation structures \parencite{hilpert_leveraging_2023}. 
As a result, calibration alone cannot fully correct for large prevalence disparities, underscoring the methodological importance of context-specific recalibration when deploying early-warning systems across institutions or populations with differing risk distributions.

\section{Discussion}

This study examined whether predictive models derived from digital learning behaviors can accurately identify students who ultimately fail or are at high risk of failing end-of-term exams early in the semester, whether such models generalize across distinct instructional contexts, and whether calibrated probability estimates more accurately reflect individual risk. In line with recommendations for explicit theory-to-feature mapping and multimodal validation, our results situate digital trace predictors within SRL constructs while acknowledging that some regulatory processes remain only partially observable in logs \parencite{MatchaEA2020, DuEA2023, BernackiEA2025}. These findings align with a growing body of evidence showing that sustained, self-regulated participation in digital learning environments---such as regular online practice, timely quiz completion, and distributed engagement---reliably predicts academic success across domains \parencite{ForsterWM2018, SchwerterDBM2022, SchwerterB2024, janson_influence_2024, Janson.etal2024, Mefferd.Bernacki2023}. By demonstrating these patterns across multiple computer science courses and institutions, the present study expands prior single-course evidence, situates these findings in demanding, theory-intensive STEM coursework, and adds a cross-context validation perspective.

Across analyses, early prediction of at-risk students proved feasible. Within two to four weeks, discrimination equaled or exceeded typical benchmarks in the literature \parencite[AUC and accuracy in the .70–.80 range, see; ][]{BernackiEA2020, ArizmendiEA2023}, demonstrating practical feasibility for early identification. For within-institution transfers with similar at-risk rates, early-semester out-of-sample performance—particularly for Elastic Net—remained comparable to prior findings \parencite{BernackiEA2020, ArizmendiEA2023, bird_is_2025}. However, generalizability was constrained when models were transferred to a different institution with a different base rate of at-risk students. Although threshold adjustments based on the target prevalence mitigated some of the discrepancies, out-of-sample prediction at the other institution was always below in-sample prediction. Together, these results highlight interconnected theoretical, methodological, and practical considerations for designing predictive models that function reliably across diverse instructional contexts, particularly when institutions seek to reuse early-warning systems rather than rebuild them for each new course. 

Theoretically, the findings demonstrate which SRL-related behavioral indicators reliably predict success while also illustrating their contextual variability. Early and distributed engagement, repeated practice, and consistent completion of incentivized tasks aligned with SRL accounts of strategic engagement and effort regulation \parencite{greene_self-regulation_2024, zimmerman_becoming_2002}. Further, incentivized practice activities were particularly strong and recurring predictors, consistent with evidence that structured and incentivized opportunities for practice relate to exam performance \parencite{ForsterWM2018, SchwerterDBM2022, schwerter_differential_2025}. Although the specific week-level predictors differed across courses, the behavioral patterns captured by these indicators can be related to processes featured in theoretical frameworks of SRL. Early and distributed task initiation can be related to temporal planning and strategic forethought in Zimmerman's (\citeyear{zimmerman_attaining_2000}) forethought phase and to Stage 2 (goal setting and planning) in Winne and Hadwin's (\citeyear{winne_studying_1998}) COPES model. Submission frequency and session regularity can be related to effort regulation and sustained engagement associated with Zimmerman's (\citeyear{zimmerman_attaining_2000}) performance phase and Stage 3 (enacting study tactics) in Winne and Hadwin (\citeyear{winne_studying_1998}). These behavioral families are afforded by digital resources supplied in courses \parencite{bernacki_clickstream_2026} and have been modeled with learning analytics under constructs including temporal regulation \parencite{knight_time_2017}, sustained effort \parencite[e.g.,][]{papamitsiou_exploring_2019}, and persistence \parencite[e.g.,][]{Glick2019}.  Importantly, LMS trace data capture observable learning actions rather than the underlying cognitive and metacognitive regulation processes themselves. Consequently, the variables used here should be interpreted as potential behavioral evidence of SRL-related activity based on the affordances of the learning resources provided \parencite{bernacki_clickstream_2026} and researchers’ interpretations of metadata about events, rather than direct measurements of SRL processes.

This divergence at the surface level but convergence at the construct level of SRL-aligned behavioral families echoes multimodal validation studies showing that clusters of digital actions map onto SRL macro-processes \parencite{BernackiEA2025, Fan2022_ValiditySRL, Yu_inpress_SRLSequences}. Taken together, these results suggest that predictive models may not generalize feature-for-feature across contexts, yet theoretically interpretable families of behaviors provide a more stable conceptual lens for interpreting digital-trace patterns across learning environments. Our findings corroborate work acknowledging the complexity of educational systems, and the need for complex modeling of behavioral phenomena to capture these patterns in learning events across instructional contexts \parencite{hilpert_leveraging_2023}.

Methodologically, the study provides one of the few systematic examinations of cross-course and cross-institution generalizability in predictive learning analytics. Despite strong in-sample performance---especially for ensemble models---out-of-sample performance declined when transferring models across contexts with different pacing structures, incentive conditions, or base rates. Elastic Net maintained above-chance transfer performance in the early semester weeks within the same institution, whereas Random Forest and XGBoost often exhibited steeper performance drops. This pattern supports concerns raised in prior work that predictive models are context-bound \parencite{mathrani_perspectives_2021, sghir_recent_2023} and that portability depends on course design and risk prevalence \parencite{MatchaEA2020, bird_is_2025}. Consequently, instructors and institutions should avoid assuming that high in-sample discrimination guarantees out-of-sample accuracy and should prioritize local validation before deployment.

The analyses also underscore the complexity of interpreting feature-importance metrics. Week-to-week variability in selected predictors, coupled with the sensitivity of importance rankings to model specifications and preprocessing decisions \parencite{Rajbahadur.etal2022, Sauerbrei.etal2020}, limits their suitability for causal inference or intervention design. SRL theory therefore remains essential for interpreting digital traces, with data-driven findings serving to refine rather than replace theory \parencite{DeMooij.etal2025a, Li.etal2025, Lan.Zhou2025}. Work on temporally contextualized feature aggregation \parencite{plumley_codesigning_2024} and interaction-dominant engagement patterns \parencite{hilpert_leveraging_2023} suggests that meso-level, temporally structured predictors may provide a promising middle ground between highly individualized traces and overly coarse global indicators.

Finally, the calibration results highlight the importance of probability accuracy for equitable decision-making. Post-hoc calibration via Platt scaling improved probability estimates within contexts sharing similar base rates but could not fully correct for strong prevalence differences across institutions. Because probability estimates influence allocation of support resources, calibration plays a central role in ensuring fairness and avoiding the amplification of inequities \parencite{slade2013learning, ferguson2017evidence, PhelpsEA2024}. These findings underscore the need for context-sensitive calibration, particularly when transferring models across courses, if early-warning systems are to provide equitable and trustworthy support for diverse student populations at scale.

\subsection{Implications, limitations, and future directions}

Students' SRL-aligned digital learning indicators offered strong predictive value within courses, yet portability depended heavily on course design and base-rate alignment—underscoring that generalizability is not a default property of predictive models but a design choice that requires intentional feature engineering and local calibration. These results therefore highlight a broader principle: effective early-warning systems must be rooted in theoretically meaningful indicators, validated in the specific contexts where they are deployed, and supported by calibration routines that ensure equitable interpretation of risk estimates. Viewed together, these considerations outline a pathway toward predictive systems that are both robust and instructionally useful.


The findings highlight that SRL provides a useful framework for interpreting digital learning behaviors but also reveal that the meaning of those behaviors is shaped by instructional context. The same trace—such as submitting work early or clustering activity near deadlines—may signal proactive regulation in one course but routine compliance in another  \parencite{MatchaEA2020}. This context dependence suggests that research should move beyond identifying which SRL processes are reflected in trace data to theorizing how course structure, pacing, and incentive mechanisms shape the manifestation and predictive value of SRL-related behaviors.



\subsubsection{Limitations}

Several considerations should be acknowledged when interpreting the findings and when situating them within the broader methodological context of trace-based learning analytics.

First, the analyses focused on LMS and practice-system traces, which provide scalable and non-intrusive observations of students' interactions with digital learning environments but limit direct observation of metacognitive monitoring and judgment; some SRL processes are only weakly observable without complementary modalities \parencite{BernackiEA2025}. This trade-off reflects a broader characteristic of trace data: they enable unobtrusive measurement of learning behaviors at scale across authentic course contexts while primarily capturing observable actions rather than the underlying cognitive and metacognitive regulation processes. 
Future work could integrate complementary modalities, such as self-report, embedded assessments, or multimodal measures, to provide a richer account of how regulatory processes unfold alongside digital engagement. Doing so will also require advances in how researchers handle the substantial missingness typical of self-report measures, particularly because imputation choices can meaningfully alter downstream feature-selection results \parencite{schroeder_interpretable_2024, gunn_how_2023, du_variable_2022, wood_how_2008, romero_which_2025}. Such approaches would be difficult to implement at the scale required for cross-course or cross-institution analyses. The exclusive reliance on LMS trace data was therefore a deliberate design choice, reflecting the study's focus on generalizability across authentic instructional contexts rather than measurement depth within a single course. 

Second, similar to all other studies that rely on digital trace data, our analysis does not include offline learning behaviors, peer interactions, and other forms of engagement that may contribute meaningfully to academic performance. This constraint reflects an inherent property of trace-based measurement approaches. Focusing on LMS traces enables scalable observation of students’ learning behaviors across courses and institutions, which is essential for research questions centered on predictive generalizability. Future research may wish to link LMS traces with broader data sources or design instruments that capture offline study strategies to help clarify how online and offline processes jointly shape learning outcomes. 

Third, although the models predicted exam outcomes well overall, the specific features that appeared most important varied from week to week and from course to course. This variability likely reflects the temporal structure of the semester and the evolving nature of students’ engagement patterns as course demands change. In this sense, the observed instability highlights a broader modeling challenge: how predictive learning analytics approaches can represent temporally dynamic learning behaviors more effectively. Students' engagement patterns naturally evolve as course demands shift over time, and week-specific predictors will therefore capture different aspects of regulatory behavior depending on when in the semester they are measured. This points to the need for modeling approaches that explicitly account for temporal dynamics, such as sequence-based models or time-structured representations of learning behaviors, which may yield more stable and generalizable predictions across instructional contexts. This instability does not undermine the overall findings, but it does limit how directly instructors can use individual feature rankings to guide intervention design. A promising direction for future research is to move beyond week-specific indicators and develop more stable, SRL-aligned representations of engagement, such as temporally structured or aggregated measures that may generalize more consistently across contexts. 

Finally, although the participating courses included several hundred students, sample size limitations precluded the use of large, independent holdout sets, which would have provided stronger evidence of model robustness, particularly for the in-sample estimates. Using cross-validation instead of holdout sets may partly explain the large discrepancies observed between in-sample and out-of-sample performance of the ensemble methods. More broadly, for questions of predictive generalizability, the number and diversity of courses may be more consequential than the number of individual students, because course design, assessment structure, and incentive systems shape the meaning of digital learning behaviors. Future research using larger datasets and a broader range of courses or instructional contexts may wish to implement fully independent test sets to provide stronger evidence on model robustness and generalizability.

\section{Conclusion}

Taken together, these findings demonstrate that SRL-aligned digital traces enable accurate early identification of at-risk students, 
while also showing that the predictive utility of such traces depends strongly on contextual factors such as course design, incentive structures, and base-rate differences. Rather than transferring individual predictors across settings, generalizability appears more likely to emerge from theoretically grounded families of behaviors---such as early task initiation, distributed engagement, and consistent effort regulation---that remain interpretable across different instructional designs. More broadly, this study therefore contributes to the growing literature on trace-based learning analytics by clarifying both the predictive value of SRL-aligned behavioral indicators and the conditions under which predictive models generalize across courses and institutions. By combining strong theoretical grounding, rigorous methodological validation, and explicit attention to fairness and interpretability, predictive learning analytics can advance toward its central promise: providing timely, data-informed, and equitable support for student success in higher education.



\printbibliography

\newpage 
\section{Supplementary Online Materials}

\appendix

\section{Methods}

\subsection{Hyperparameter tuning}

Hyperparameter tuning was conducted in a 10-fold cross-validation to ensure robustness and account for variability in at-risk prevalence. Fold sampling was stratified, maintaining comparable proportions of at-risk and non–at-risk students within each fold. Hyperparameters were optimized via grid search, selecting configurations that maximized the area under the receiver operating characteristic curve \parencite[AUC; ][]{friedman_elements_2001}. The AUC metric was chosen because it does not depend on a fixed probability threshold and is less sensitive to class imbalance compared to accuracy or deviance \parencite{sun2009classification}. Given the relatively small sample sizes, models with hyperparameters optimized in the cross-validation, were trained and evaluated on the complete dataset rather than separate training and test partitions. Subsequent analyses then focused on testing these trained models with data from a different course or institution to evaluate out-of-sample generalizability.

For Elastic Net, we varied the mixing parameter $\alpha$ from {0, 0.1, ..., 1} and the regularization strength $\lambda$ on a grid of length $100$ from $0.001$ to ${1,000.00}$ with log-equidistant steps. 
For Random Forest, the number of features sampled per split (\textit{mtry}) varied around the default value of the square root of $p$, where $p$ is the number of predictors \parencite{WrightZ2017}. Accordingyl, we varied \textit{mtry} on the integers between the default minus half the default and the default plus half the default. To evaluate larger values we added default times 2.5 and default times 3 to the tuning grid. The minimum node size was set to one of the following values: 10, 12, 15, 17, 20, 23, 25, or 30. The number of trees was set to one of the following values: 500, 1000, or 2000. For XGBoost, the maximum tree depth was varied on $3, 4, \ldots, 10$ and the minimum child weight was between 1 and 10. The subsample ratio was between 0.5 and 1 and the column sampling ratio was between 0.5 and 1.  Each range was discretized into eight candidate values.

\subsection{Classification Thresholding, Evaluation Metrics, and Feature Importance}

All three methods generate predicted class probabilities which were converted into binary classifications by dichotomizing at a threshold. The thresholds were found by maximizing the Youden index \parencite{youden1950index}, which is the sum of sensitivity and specificity minus one, to strike a balance between false positives and false negatives \parencite{schisterman2005optimal}. We report AUC as a metric independent of the threshold. Additionally, we report accuracy,  sensitivity (recall), specificity, $F_1$ score and Cohens'~$\kappa$, which all depend on the threshold. The $F_1$ score takes into account both, sensitivity and specificity \parencite{powers2011evaluation}, and Cohens'~$\kappa$ can be interpreted as chance-corrected agreement of prediction and ground truth \parencite{ArizmendiEA2023}. Feature importance was assessed via standardized coefficients (Elastic Net), Gini impurity (RF), and out-of-bag permutation importance (XGB) within each week.

\subsection{Model Quality}

Accuracy measures how often the model is correct and can be calculated by dividing the total number of correct classifications by the total number of observations:
\begin{equation}
\text{accuracy} \;=\; \frac{\text{total number of correct classifications}}{\text{total number of observations}},
\end{equation}
where higher values indicate a greater overall rate of correct predictions. 

Sensitivity reflects the model’s effectiveness in identifying successful students (true positives) and is calculated as
\begin{equation}
\text{sensitivity} \;=\; \frac{\text{number of true positive decisions}}{\text{number of positive decisions}},
\end{equation}
where a higher value implies that more successful students are correctly classified. 

Specificity captures the model’s ability to correctly categorize unsuccessful students (true negatives) and can be expressed by
\begin{equation}
\text{specificity} \;=\; 1 \;-\; \frac{\text{number of false positive decisions}}{\text{number of negative decisions}},
\end{equation}
so higher specificity means that fewer unsuccessful students are misclassified as successful. 

AUC quantifies the relationship between sensitivity and specificity across different classification thresholds and ranges from 0 to 1, where 0.5 indicates no discriminative power (i.e., random guessing) and 1 corresponds to perfect classification. 
AUC is commonly defined as the \textit{Area under the Receiver Operating Characteristic} (ROC) curve, which plots the True Positive Rate (TPR) against the False Positive Rate (FPR) at various classification thresholds. In continuous form, it can be written as:
\begin{equation}
\text{AUC} \;=\; \int_0^1 \text{TPR}\bigl(\text{FPR}^{-1}(x)\bigr)\,dx,
\end{equation}
where \( x \) represents all possible false positive rate values from 0 to 1.
In practice, one often uses a discrete approximation, such as the trapezoidal rule applied to the empirical ROC curve. Alternatively, when the model outputs probability scores \(\hat{p}_i\) for each instance \(i\), another well-known formula relies on ranks:
\begin{equation} 
\text{AUC} \;=\; \frac{1}{n_+\,n_-}
\sum_{i \in \{\text{positives}\}} \sum_{j \in \{\text{negatives}\}}
\mathbf{1}\Bigl(\hat{p}_i > \hat{p}_j\Bigr),
\end{equation}
where \(n_+\) is the total number of positive instances (e.g., successful students), \(n_-\) is the total number of negative instances (e.g., unsuccessful students), and \(\mathbf{1}(\cdot)\) is an indicator function that takes the value of 1 if \(\hat{p}_i > \hat{p}_j\) and 0 otherwise. Essentially, this formula counts the proportion of all positive–negative pairs in which the positive instance has a higher predicted score than the negative instance. Consequently, a higher AUC signifies stronger performance in distinguishing between students who succeed and those who do not. 

\subsection{Calibrated Probabilities} 

Although metrics such as accuracy, sensitivity, specificity and AUC provide valuable insight into a model's ability to discriminate between classes (e.g. pass vs.\ fail), they do not necessarily indicate whether the predicted probabilities themselves are correctly scaled. In other words, models may produce probability estimates that are systematically biased, either too high or too low, thereby misrepresenting the true likelihood of an event. This concern is particularly important when decisions or interventions depend on the likelihood that a student will pass an exam, rather than simply assigning each student to a pass/fail category.

Calibration refers to the extent to which a model's predicted probabilities match the actual observed frequencies. For example, if a model predicts an 80\% pass probability for a group of students, then close to 80\% of those students should actually pass, provided the model is well calibrated. Models such as Elastic Net, Random Forest and XGBoost predict probabilities denoted by \(\hat{p}\). For example, in Lasso (easier version of Elastic Net), if we let \(\hat\beta_0, \hat\beta_1, \ldots, \hat\beta_k\) be the regularised regression coefficients, then the predicted probability of passing given the predictors \(x_1, \dots, x_k\) is estimated by:
\begin{equation}
\hat{p}\,(x_1, \ldots, x_k) \;=\;
\frac{\exp(\hat\beta_0 \;+\; \hat\beta_1 x_1 \;+\; \dots \;+\; \hat\beta_k x_k)}{1 \;+\; \exp(\hat\beta_0 \;+\; \hat\beta_1 x_1 \;+\; \dots \;+\; \hat\beta_k x_k)}.
\end{equation}
These probabilities, however, can deviate from actual frequencies if the model is miscalibrated. A straightforward method to evaluate calibration is to bin the predicted probabilities and compare them to observed proportions in each bin.

If the predicted probabilities are found to be miscalibrated, they can be recalibrated by applying a monotonically increasing function to \(\hat{p}\). A common approach is Platt scaling \parencite{PlattEA1999}, which adjusts the probabilities to better match the true outcomes. It is crucial to evaluate the performance of any calibration method on data not used for training (i.e. out-of-sample) to ensure that the adjustments generalize beyond the initial sample.

In-sample calibration involves training and applying Platt scaling on the same dataset (e.g., Course Logic-A), whereas out-of-sample calibration trains the calibration model on one course’s data (again, Course Logic-A) and applies it to another course (e.g., Course TCS-A or Course Logic-B). By examining the degree to which these recalibrated probabilities match observed outcomes in a new context, we verify not only the model’s predictive power but also the reliability of its probability estimates for guiding decisions in educational settings. In this way, calibration of probabilities increases the utility of predictive models in educational contexts by providing accurate assessments of each student's likelihood of success.

%
%



\subsection{Overview generated variables
}

\begin{table}[htbp]
\centering
\caption{Complete list of generated variables prior to omission for multicollinearity}\label{tab_fulllist}
\begin{tabular}{ll}
\hline
\textbf{Abbreviation} & \textbf{Description} \\
\hline
P   & Paper-based practice \\
I   & Incentivized digital practice \\
NI  & Non-incentivized digital practice \\
\hline
Mean & Mean over all weeks \\
SD   & Standard deviation over all weeks \\
\hline
redu1 & Tasks of before the deadline \\
redu2 & Tasks of after the deadline \\
\hline
eng1  & Number of solution deliveries \\
eng2  & Time to deadline (in days) \\
eng3  & Indicator: submission took place on the last day \\
eng4  & Indicator: all submissions were made before the last day \\
eng5  & Time on page (in minutes) \\
eng6  & Indicator: any interaction before the last day \\
eng7  & Indicator: any interaction after successful submission \\
eng8  & Time on page after successful submission (in minutes) \\
eng9  & Number of weekdays worked on tasks \\
eng10 & Number of sessions per week/fortnight \\
\hline
per1 & Proportion of correctly completed pages \\
per2 & Proportion of correctly completed tasks \\
per3 & Proportion of correctly completed tasks on started pages \\
per4 & Weighting of correctly completed pages (LMS definition) \\
\hline
par1 & Proportion of tasks with submission \\
par2 & Proportion of pages with submission \\
\hline
\end{tabular}
\end{table}

%
%

\section{Results}

\subsection{Using Course TCS-A and Logic-B as reference}

For Elastic Net, out-of-sample AUCs improved slightly when Course TCS-A served as the reference course: predictions for Course Logic-A ranged from approximately .63–.77  and for Course Logic-B from .63–.78. By comparison, when Course Logic-A was the reference (baseline in RQ1b), AUCs targeting Course TCS-A started around .70–.80 (Weeks 1–4) dropped mid-semester to between .53–.55 mid-semester, and recovered to between .71-.73, while AUCs targeting Course Logic-B remained between .50–.64. Using Course Logic-B as the reference yielded Elastic Net AUCs of between .66–.74 for Course TCS-A and between .61–.71 for Course Logic-A, again slightly above the baseline values observed when Course Logic-A was the reference.

By contrast, Random Forest and XGBoost continued to exhibit weak cross-course generalization under source-threshold transfer, regardless of the reference course. AUCs clustered between $\approx$.58 and .75, frequently accompanied by degenerate operating points (e.g., specificity = 0 or sensitivity = 1.00), leading to F$_1$ = 0 or undefined (NaN) and Cohen's $\kappa$ $\approx$ 0---a pattern consistent with earlier results using Course Logic-A as reference. Thus, shifting the reference course produced small gains in out-of-sample discrimination for Elastic Net but did not mitigate the threshold-transfer failures that continued to generate imbalanced predictions for the ensemble models.

\begin{figure}[htb]
  \centering
  \caption{Weekly model quality metrics for Elastic Net, XGBoost, and Random Forest --- Course TCS-A as reference, progressive feature aggregation, source-threshold transfer}
  \label{fig:gesamt_ref_course1_os1_course2_os2_course3}
  \includegraphics[width=0.75\linewidth]{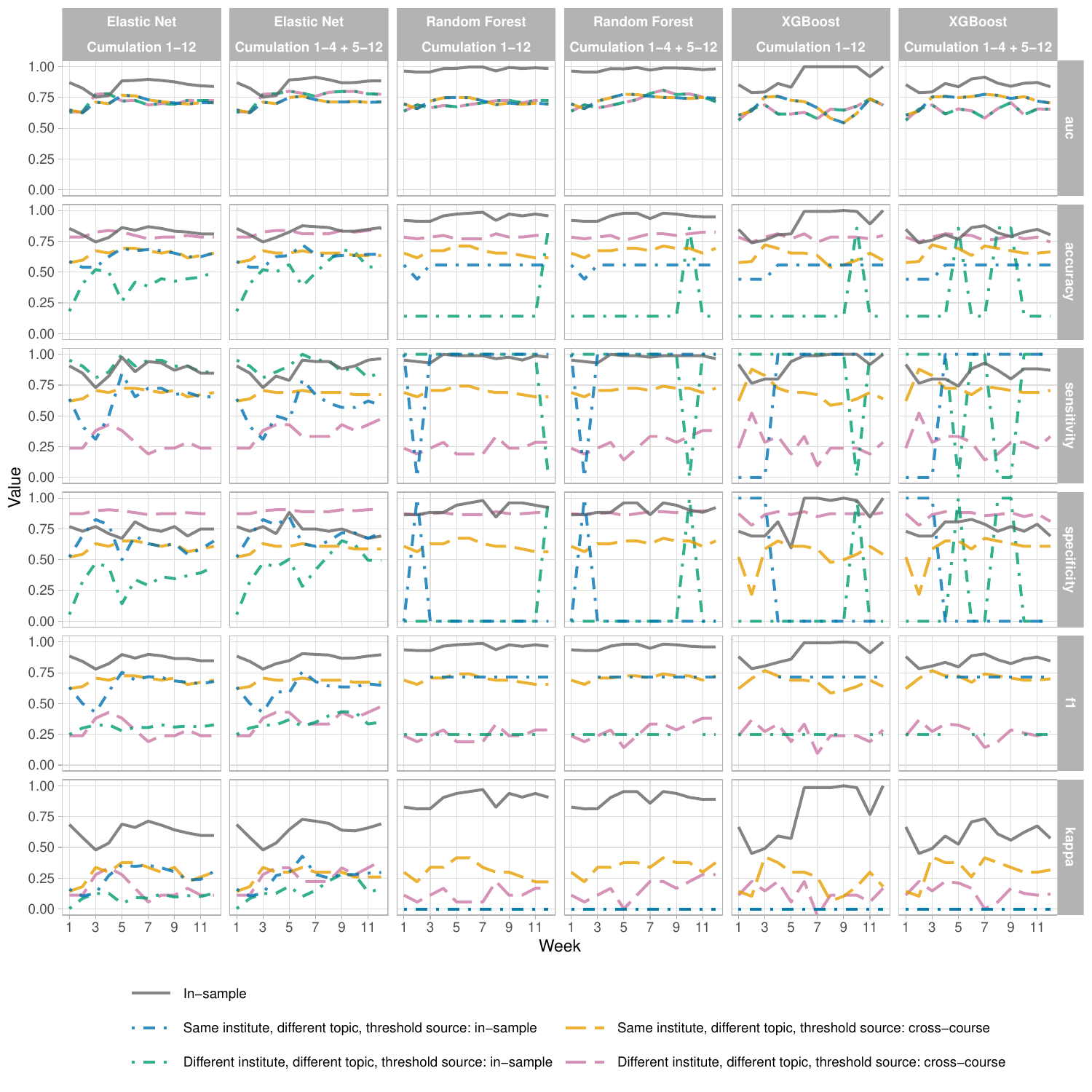}  
  \vspace{0.5em}
\end{figure}

\begin{figure}[htbp]
  \centering
  \caption{Weekly model quality metrics for Elastic Net, Random Forest, and XGBoost---Course Logic-B as reference, progressive feature aggregation, source-threshold transfer}
  \label{fig:gesamt_ref_course3_os1_course1_os2_course2}
  \includegraphics[width=0.8\linewidth]{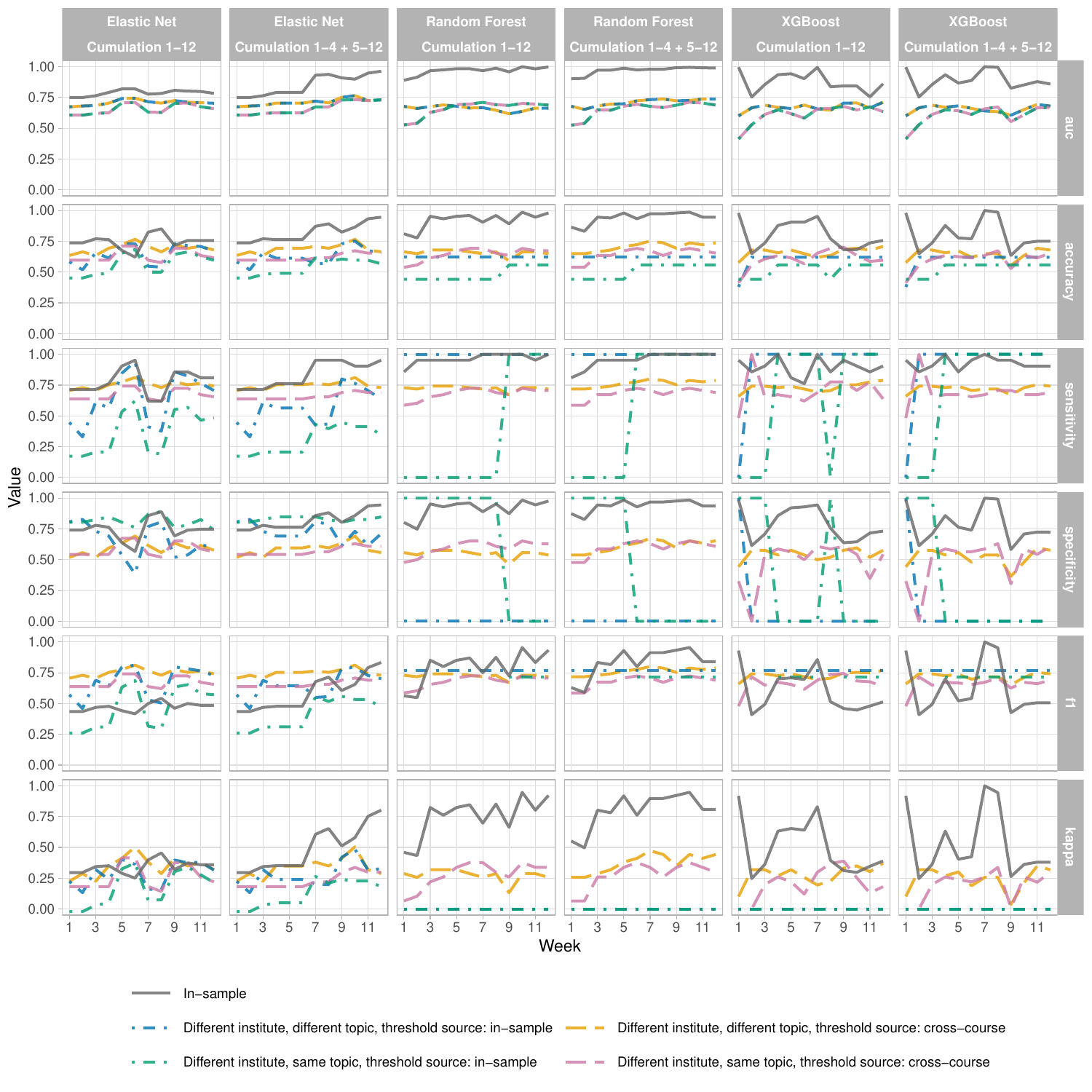}  
  \vspace{0.5em}
\end{figure}

\begin{figure}[htbp]
  \centering
  \caption{Variable importance/selection across weeks and courses -- Elastic Net Course Logic-A}
  \label{fig:importance_en2}
  \includegraphics[width=\linewidth, page = 2]{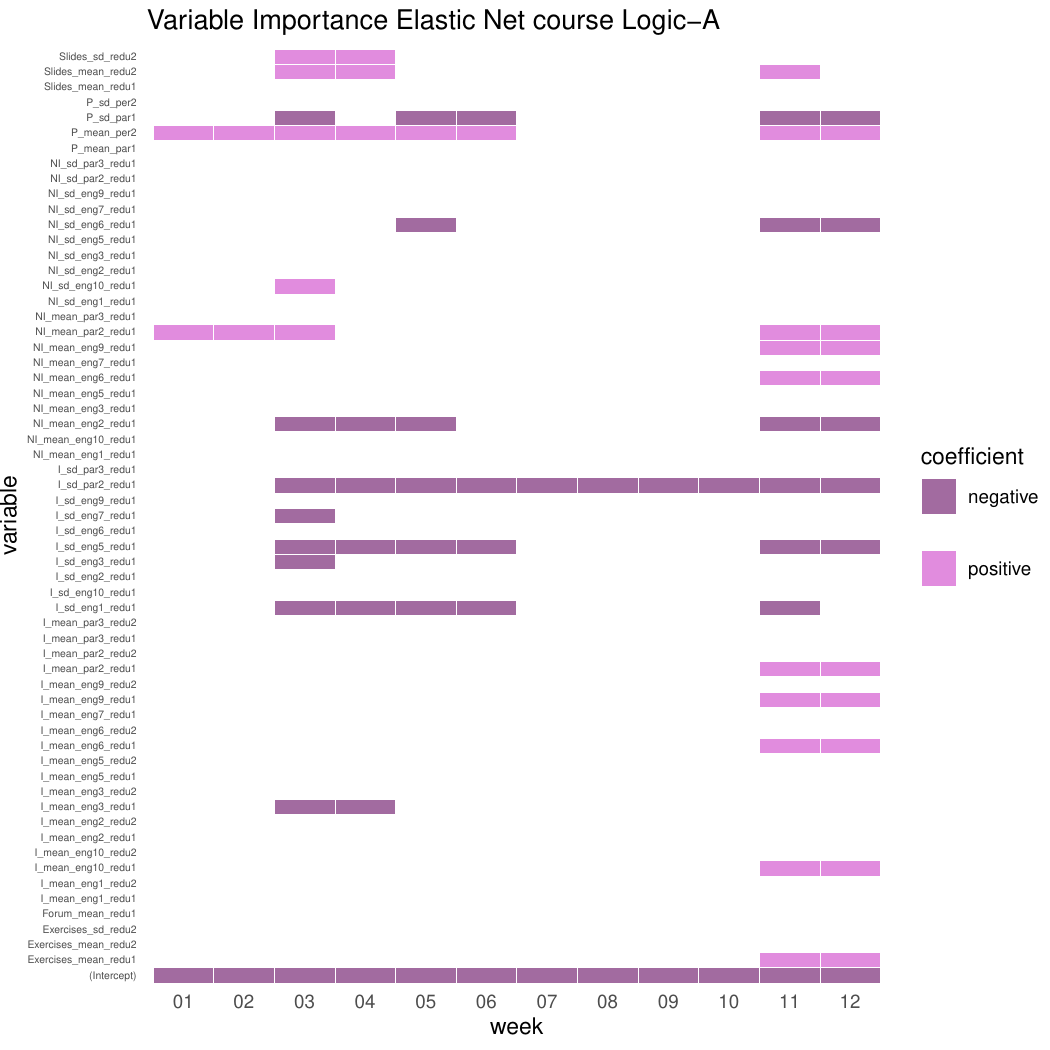}
  \vspace{0.5em}
\end{figure}

\subsection{Variable Importance for Random Forest and XGBoost}

\begin{figure}[htbp]
  \centering
    \caption{Variable importance/selection across weeks and courses -- Random Forest all courses}
  \label{fig:importance_rf_all}
\includegraphics[width=\linewidth, page = 12]{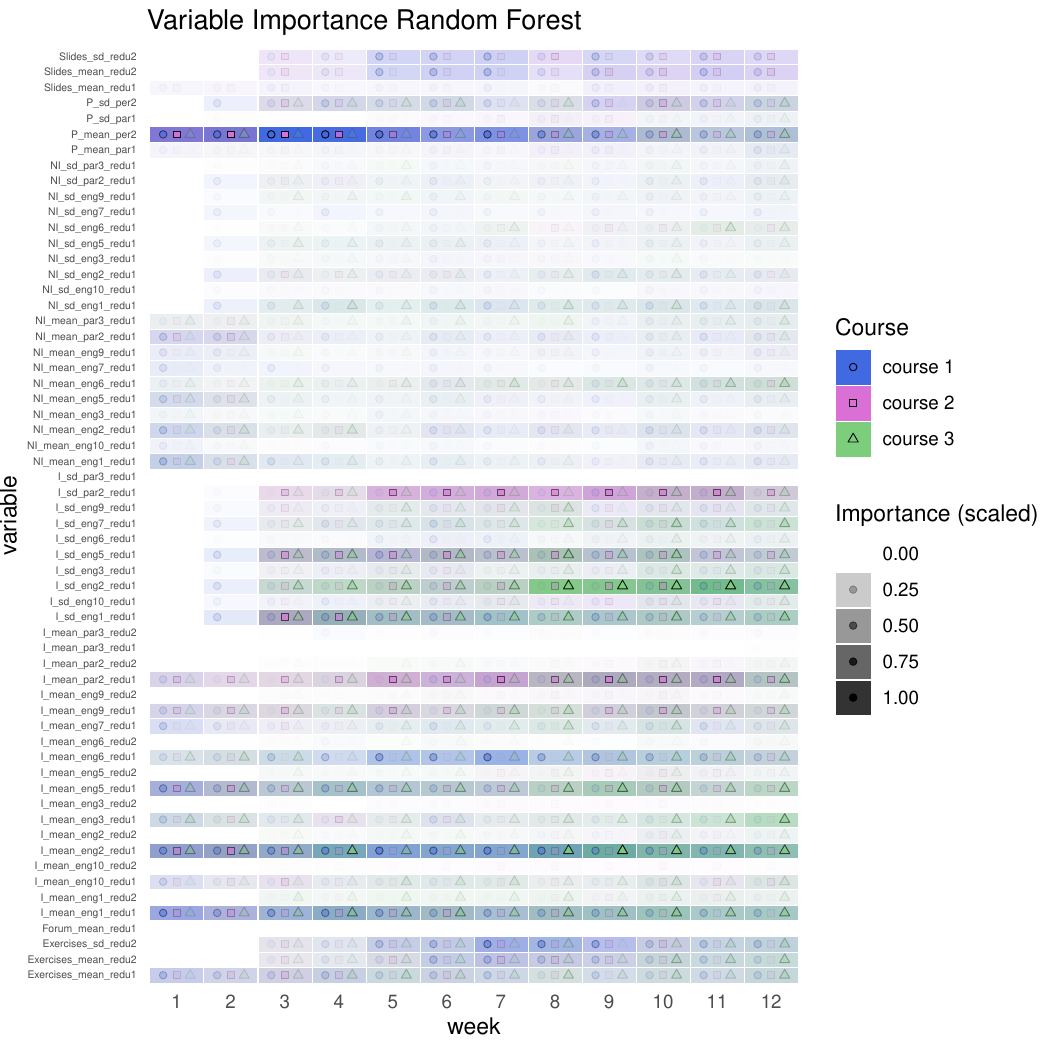}
\vspace{0.5em}
\end{figure}

\begin{figure}[htbp]
  \centering
    \caption{Variable importance/selection across weeks and courses -- XGBoost all courses}
  \label{fig:importance_xgb_all}
\includegraphics[width=\linewidth, page = 8]{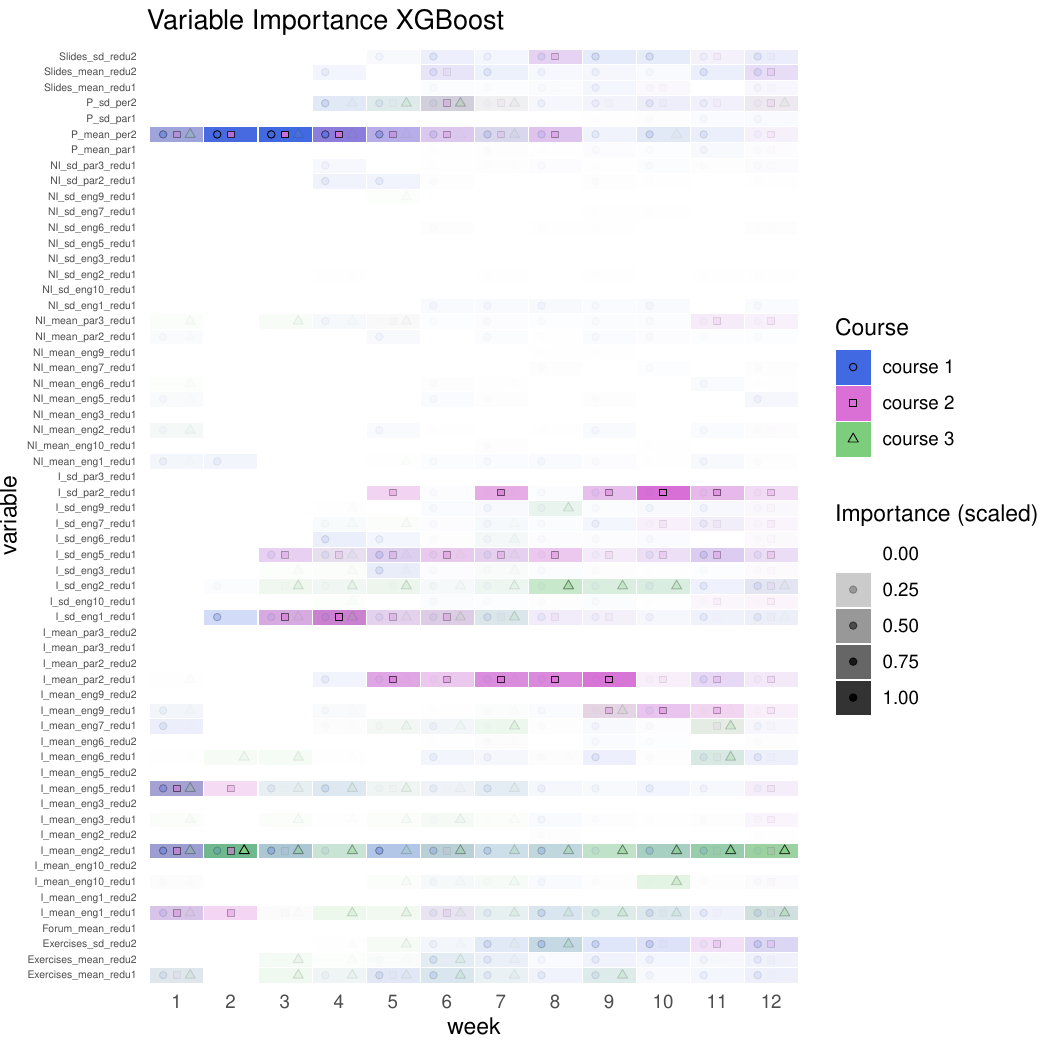}
\vspace{0.5em}
\end{figure}

\subsection{Calibration plots for Random Forest and XGBoost}

\begin{figure}[htbp]
  \centering
  \caption{Calibration plots (Week 3--6, Random Forest)}
  \label{fig:calibration_rf}
  \includegraphics[width=\linewidth]{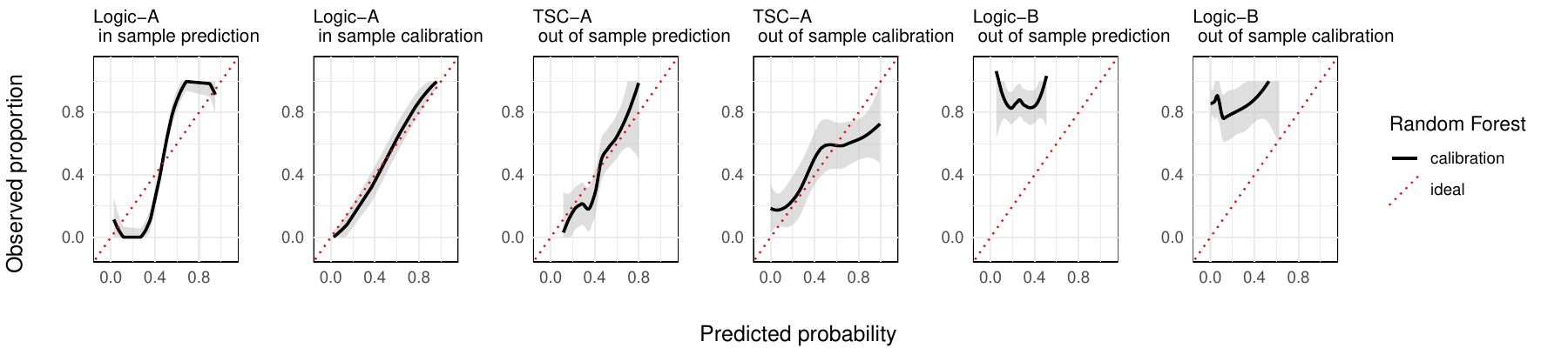}\\
  \includegraphics[width=\linewidth]{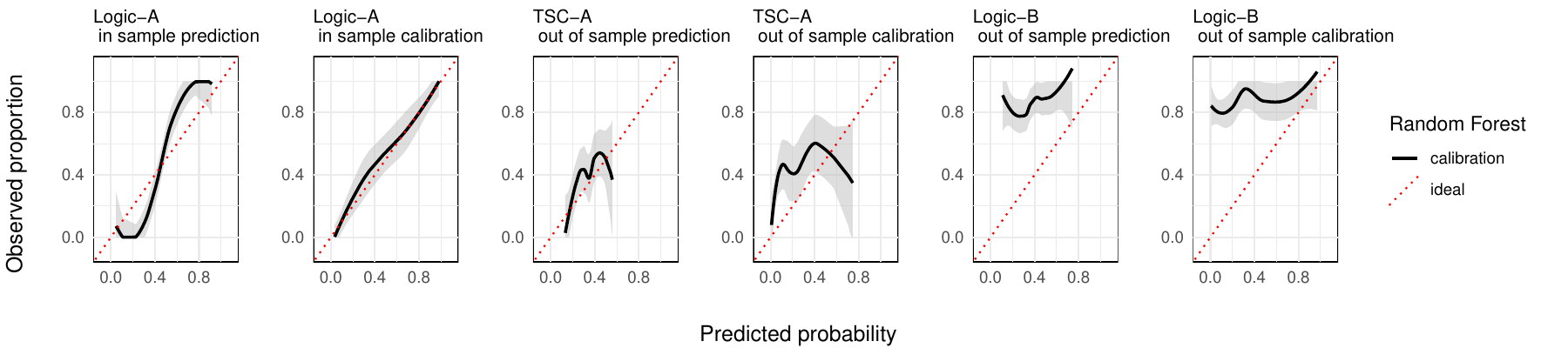}\\
  \vspace{0.5em}
\end{figure}

\begin{figure}[htbp]
  \centering
  \caption{Calibration plots (Week 3--6, XGBoost)}
  \label{fig:calibration_xg}
  \includegraphics[width=\linewidth]{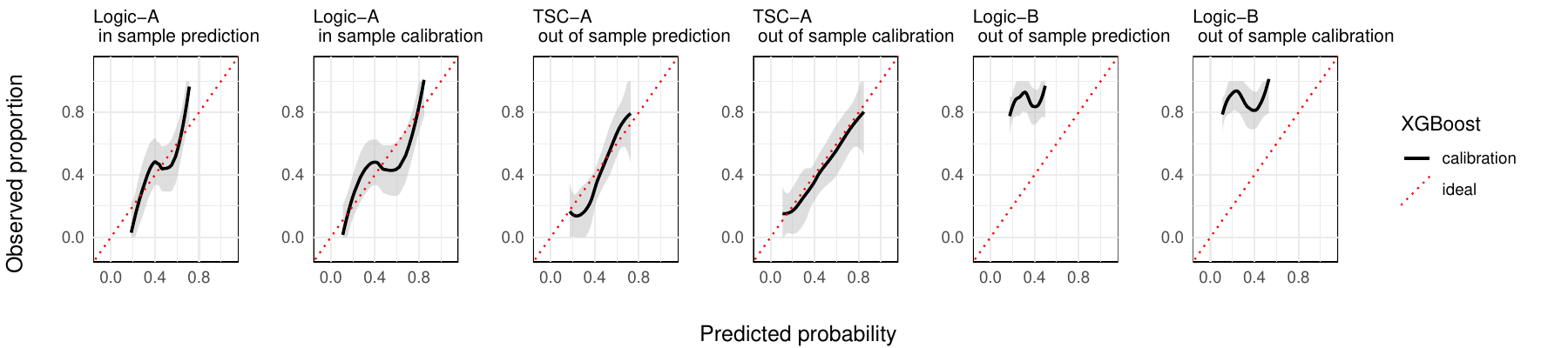}\\
  \includegraphics[width=\linewidth]{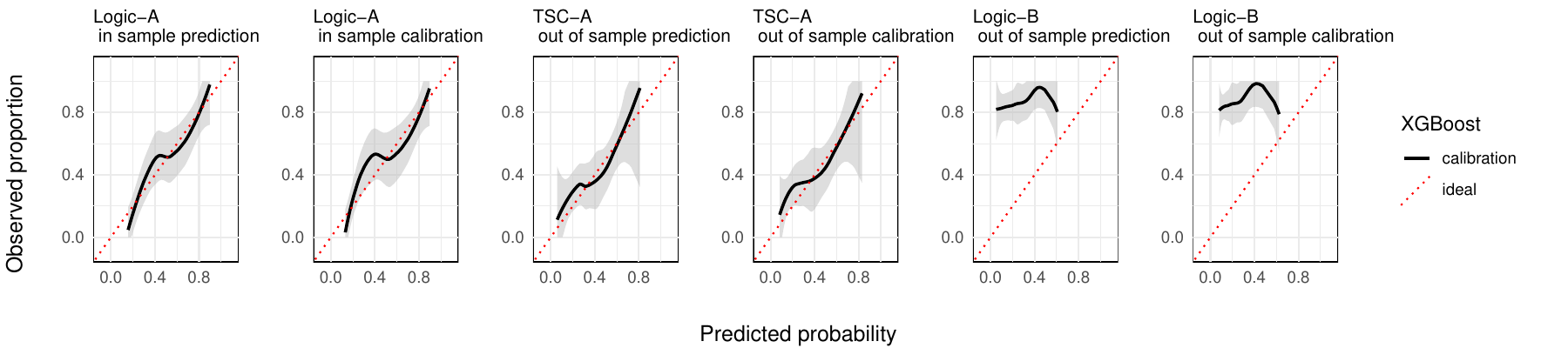}\\
  \vspace{0.5em}
\end{figure}
 
\end{document}